\documentclass[10pt]{article}
\usepackage{graphicx}% Include figure files

\usepackage{epsfig}
\usepackage{latexsym}
\usepackage{amsmath}
\usepackage{eurosym}

\begin{document}

\pdfoutput=1
%\begin{frontmatter}

\title{Simple approaches on how to discover  promising strategies for efficient  enterprise performance,\\  at time of crisis in the case of SMEs : \\   Voronoi clustering   and outlier effects  perspective }

\author{ Marcel Ausloos \textsuperscript{$a,b,c$} \\
 Francesca Bartolacci \textsuperscript{$d$} \\
 Nicola G. Castellano \textsuperscript{$e$} \\
 Roy Cerqueti \textsuperscript{$f$}\\ \\
 $^{a}$ School of
Business, University of Leicester. \\Brookfield, Leicester, LE2
1RQ, United Kingdom.\\ Email: ma683@le.ac.uk \\$^{b}$
Department of Statistics and Econometrics, \\Bucharest University of Economic Studies, \\ Calea Dorobantilor 15-17, Bucharest, 010552 Sector 1, Romania. \\
  Email: marcel.ausloos@ase.ro
\\$^{c}$
GRAPES. rue de la Belle Jardiniere, 483/0021,
\\B-4031, Liege Angleur, Belgium. \\Email:
marcel.ausloos@ulg.ac.be
\\ $^{d,f}$ Department of Economics and Law, \\University of
Macerata. Via Crescimbeni, 20, I-62100, Macerata, Italy.
 \\ $^e$ Department of Economics and Management,\\ University of Pisa.
Via Ridolfi 10, I-56124, Pisa, Italy.
\\ 
 $^{d}$ Email:  bartolacci@unimc.it\\$^{e}$ Email: nicola.castellano@unipi.it\\
   $^{f}$ Email:  roy.cerqueti@unimc.it}
%\date{\today}
\maketitle
\newpage
\begin{abstract}

This paper analyzes the connection between innovation activities of
companies -- implemented before a financial crisis -- and their
performance -- measured after such a time of crisis. Pertinent data
about companies listed in the STAR Market Segment of the Italian
Stock Exchange is analyzed. Innovation is measured through the level
of investments in total tangible and intangible fixed assets in
2006-2007, while performance is captured through growth -- expressed
by variations of sales or of total assets,  -- profitability --
through ROI or ROS evolution, - and productivity -- through asset
turnover or sales/employee in the period 2008-2010. The variables of
interest are analyzed and compared through statistical techniques
and by adopting a cluster analysis. In particular, a Voronoi
tessellation is implemented in a varying centroids framework. In
accord with a large part of the literature, we find that the
behavior of the performance of the companies is not univocal when
they innovate. The statistical outliers are the best cases in order
to suggest efficient strategies. In brief, it is found that a
positive rate of investments is preferable.
\end{abstract}

\vskip0.2cm

%Keywords:  Innovation, business performance, financial statements, STAR Market, cluster analysis, Voronoi tessellation.

\section{Introduction} \label{Introduction}

This chapter is based on three recent papers :
\begin{itemize}
\item
(Bartolacci et al., 2015) F. Bartolacci, N.G. Castellano, and R. Cerqueti (2015). The impact of innovation on companies' performance: an entropy-based analysis of the STAR Market Segment of Italian Stock Exchange, Technology Analysis and Strategic Management 27, 102-123.
\item
 (Ausloos et al., 2018a)
M. Ausloos, F. Bartolacci, N.G. Castellano, and R. Cerqueti (2018). Exploring how innovation strategies at time of crisis influence performance: a cluster analysis perspective, Technology Analysis and Strategic Management 30, 484-497.
\item
 (Ausloos et al., 2018b) M. Ausloos, R. Cerqueti,   F. Bartolacci,  and  N. G Castellano (2018).
 SME investment best strategies. Outliers for assessing how to optimize performance,
 Physica A 509, 754-765.
 \end{itemize}

The connection between innovation strategies (usually taken as the investments) of companies
if implemented before a financial crisis  and their performance  measured after crisis time are interesting aspects of  small and medium size enterprises (SME) economic life.

In fact, Latham and Braun (2011), in {\it "Economic recessions, strategy, and performance: a synthesis"}
%Journal of Strategy and Management, 4 (2), 96-115  (2011):
claimed that

\begin{center}
% \indent
{\it "Despite the episodic pervasiveness of recessions and their    destructive     impact on firms, a void exists in the management literature examining   the intersection between recessions, strategy, and performance”}.
\end{center}

Therefore, it seems worthwhile to reflect on such  connections
considering practical cases. Thus, we have  considered companies
listed in the STAR Market Segment of the Italian Stock Exchange in
recent times. SME innovation is here below measured through the
level of investments in total tangible and intangible fixed assets
in [2006-2007], while performance is captured through (i) growth --
expressed by variations of sales (DS) and variations of total assets
(DA),  -- (ii) profitability -- through returns on investments (ROI)
and  returns on sales (ROS), -- and (iii) productivity -- through
asset turnover (ATO) or sales per employee (S/E) in the period
[2008-2010].

%The variables of interest are analyzed and compared through statistical techniques and by adopting cluster analysis. In particular, a Voronoi tessellation is also implemented in a varying centroids framework. In accord with a large part of the literature,we find that the behavior of the performance of the companies is not univocal when they innovate. The statistical outliers are best to suggest efficient strategies.

In the Milano STAR market, 71 companies of mid-size  are listed, at
the time of study: their capitalization value was  about between 40 million
and 1 billion euros.  Since their activity and innovation levels are
different from "industrial companies", whence since their
performance should be measured in a different way, we have removed
banks and insurance institutes  from our analysis. Thus, in the
following, the segment is reduced to 62 SMEs\footnote{The  below
displayed data can be obtained from the authors upon  request  as
Excel tables.}. For completeness, the 62 SMEs at the time of study are given in  Table \ref{STARnames}.

We discuss a formal  method, based on Voronoi tessellation (Voronoi,
1908), yet we depart from the original formulation of Voronoi by
introducing a concept of weighted Euclidean distances, hence leading
to asymmetry (see formulas (\ref{dalpha}) and (\ref{dbeta}) below).
In our approach, we a priori define some reference points - so
called "centroids", each centroid identifying a cluster whose
elements are at a distance smaller than that to the other centroids.

For more information, let us mention that the use of Voronoi
tessellation  can be found in Liu et al. (2009), Yushimito et al.
(2012) and Vaz et al. (2014).

% clustering techniques are largely employed to analyse the performance at country, industrial district or firm level (see e.g. Zahra and Covin, 1994; Gligor and Ausloos, 2007, 2008a, 2008b).

Such a cluster analysis is employed to investigate the determinants
of innovation and innovation-performance focused on a single
industry (Tseng et al., 2008),
% a single country (Dwyer and Mellor, 1993; Vaz et al., 2014; Agostini et al., 2015),
or  on different industries   (Pavitt, 1984; Cesaratto and Mangano, 1993; Leiponen and Drejer, 2007).

 \section{Data} \label{Data}
A few notations are to be introduced   for  easy readability of the
following tables and figures:

\begin{itemize}
\item TIAXyy represents the level of total intangible assets (excluding goodwill) in year 20yy;
\item TTAyy is the level of total tangible assets (excluding properties) in 20yy;
\item DSyy stands for sales variations in year 20yy
\item DAyy is total assets variations, in year 20yy
%\item DLabyy means employees variations, in year 20yy
\item ROIyy means the return on investments  in year 20yy
\item ROSyy  means the return on sales in year 20yy
\item ATOyy represents asset turnover, in year 20yy
\item S/Eyy stands for sales per employee, in year 20yy
\end{itemize}

\begin{itemize}
\item
 the lowest TTA  value is called TTA1, while the highest TTA  is TTA2
\item  their average is : $< $TTA$>$2 = (1/2) (TTA1 + TTA2)
 \end{itemize}
 which  in fact, due to the time interval of interest,  is equal to  (TTA06 + TTA07). Similarly,

\begin{itemize}
\item $<$TIAX$>$2 is the average total intangible asset (excluding goodwill) over 2 years: [2006-2007];
"obviously",   $< $TIAX$>$2 = (1/2) (TIAX1 + TIAX2) = (1/2) (TIAX06 + TIAX07).
%\item $<$TTA$>$2 represents the average of the total tangible assets (excluding properties) over 2 years: [2006-2007];
 \end{itemize}

\begin{figure}   %1
    \begin{center}%centering
\includegraphics[scale=.230] {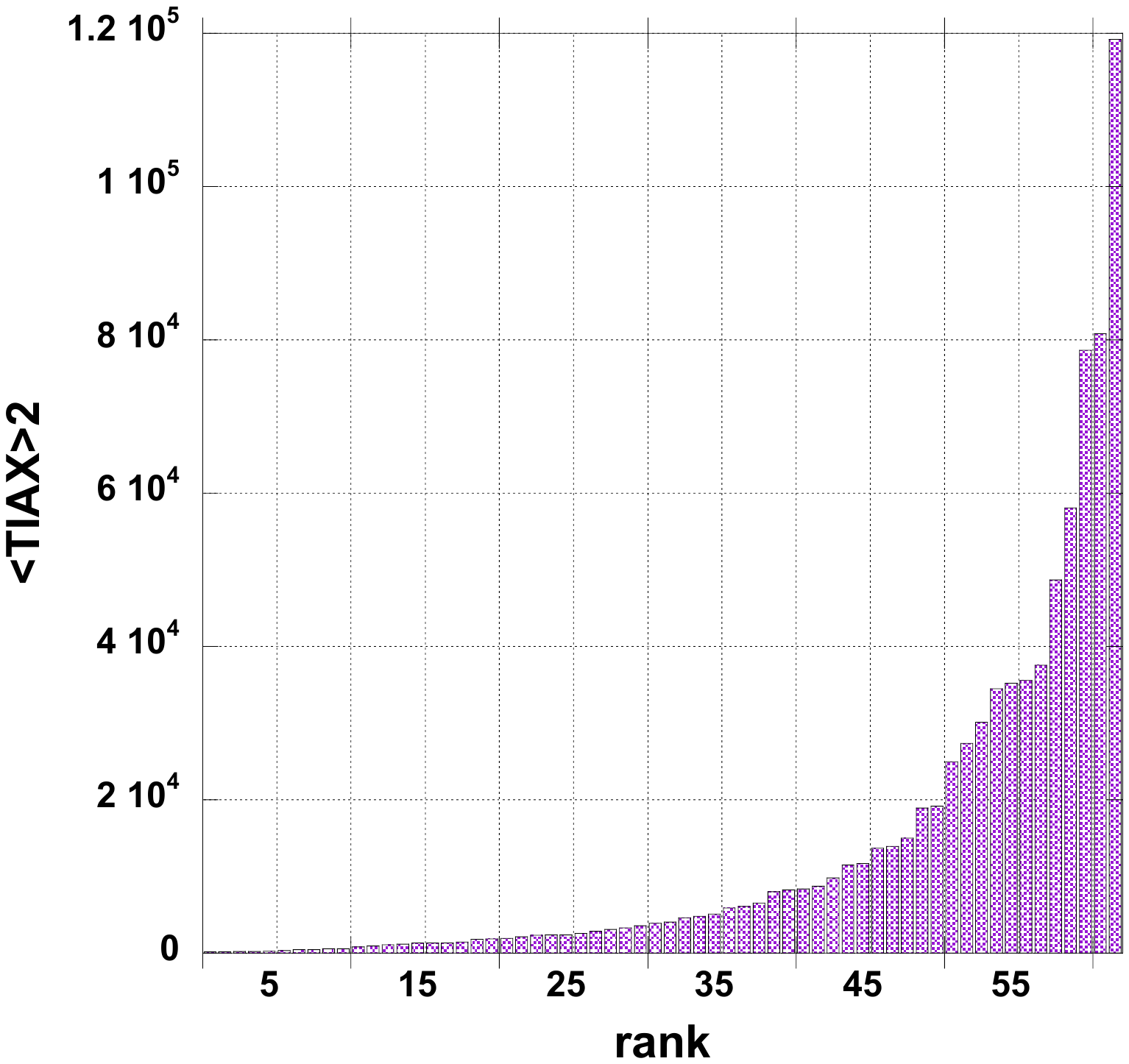}
~
\includegraphics[scale=.230] {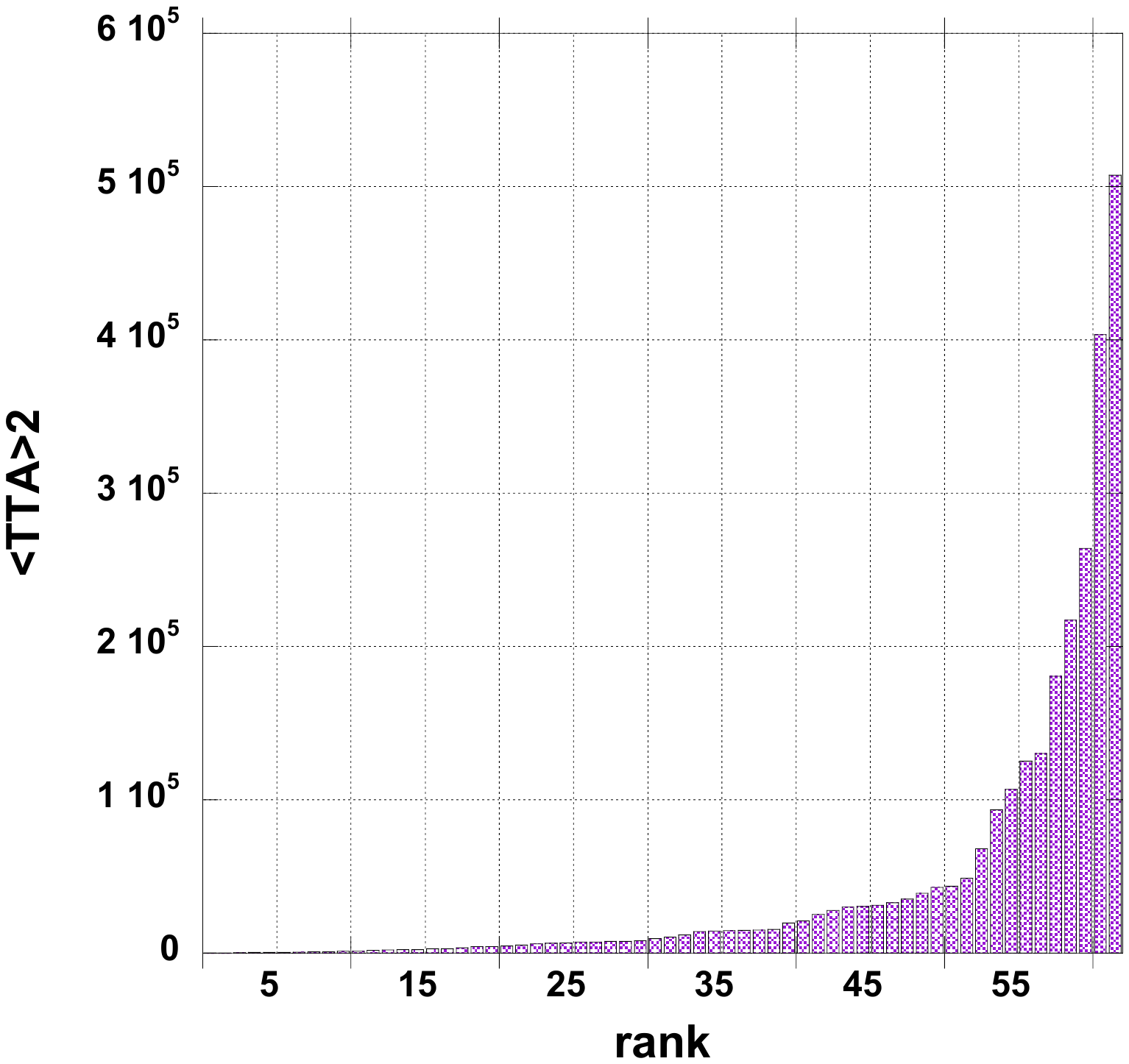}
%\label{Plot6TITA2ranked}
\caption{  Left panel:  $<$TIAX$>$2. Right panel: $<$TTA$>$2; thus each averaged over 2 years: [2006-2007]; both data are  ranked  in increasing order, - for the 62 SMEs discussed in the text. }
\label{Plot6TIAX2TTA2ranked}
\end{center}
\end{figure}

\begin{figure}[t]
               \includegraphics[width=0.6\textwidth, height=0.955\textwidth]
            {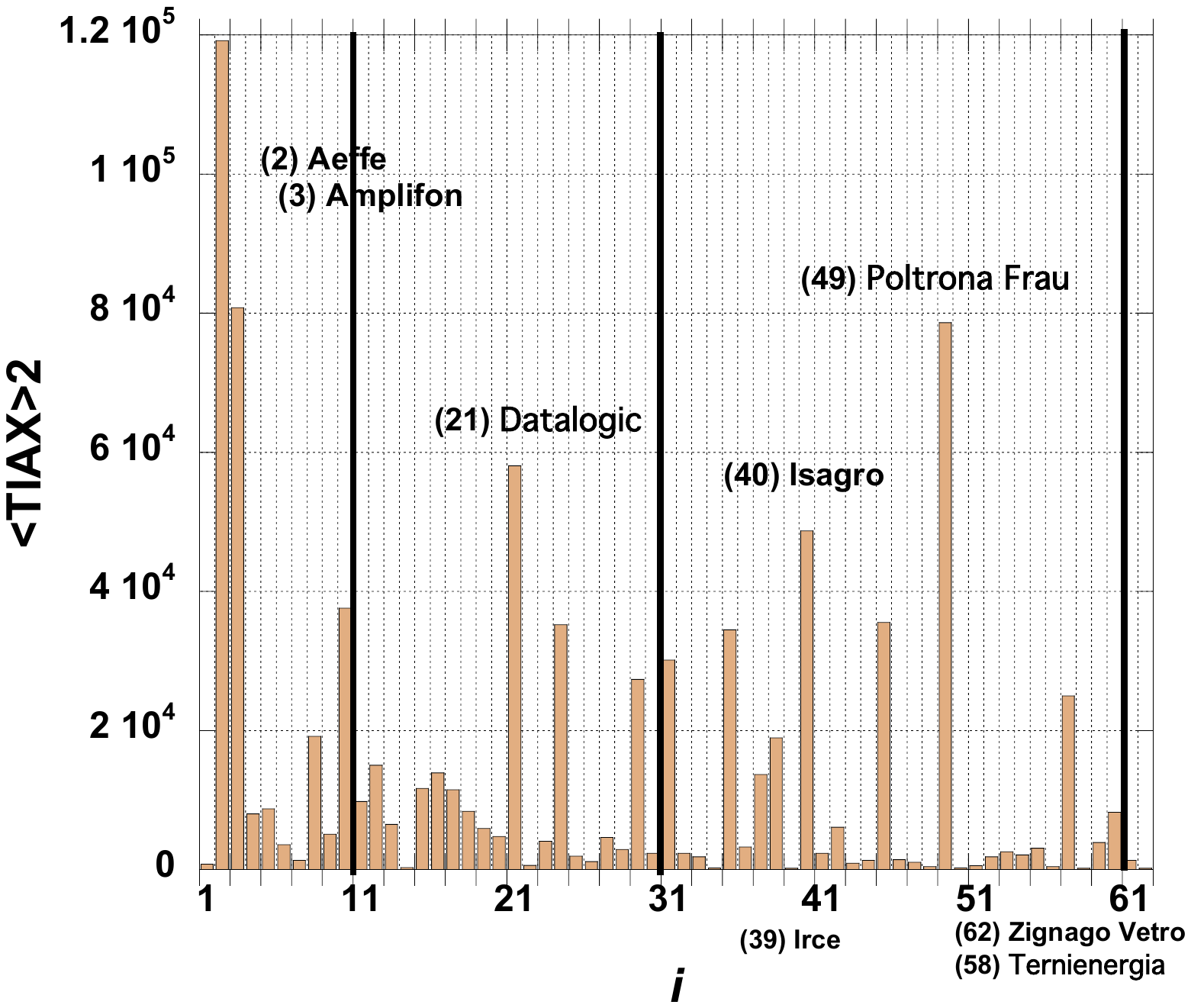}
            \label{fig:Plot12TIAX2}
      %  ~
            \includegraphics[width=0.6\textwidth, height=0.95\textwidth]
            {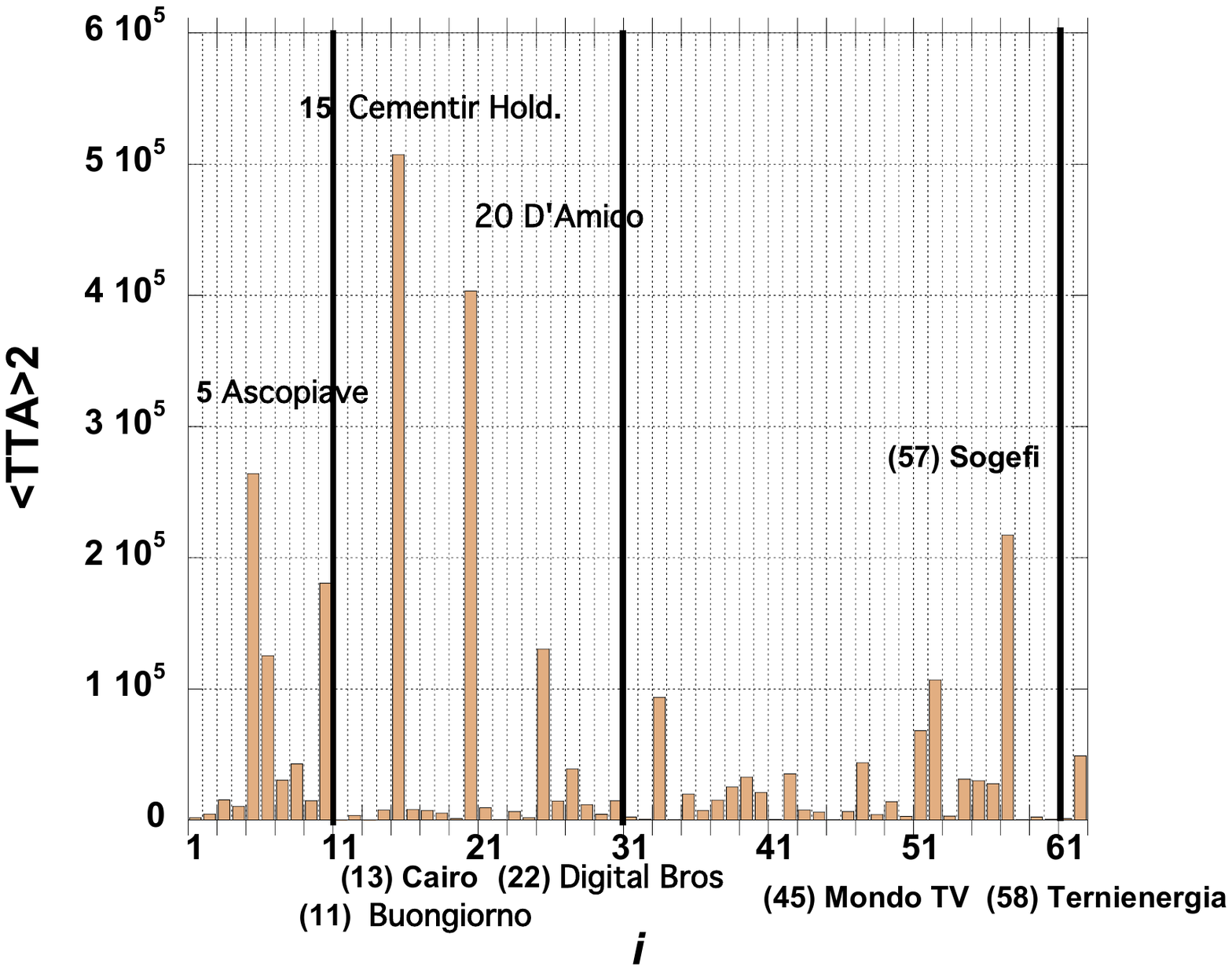}
           \label{fig:Plot11TTA2}
         \caption{(color online)    Left panel: $<$TIAX$>$2.          Right panel:  $< $TTA$>$2;  thus each averaged over 2 years: [2006-2007],  - for the 62 SMEs, ranked in alphabetical order as in Table \ref{STARnames},
         particularly pointing to  a few relevant SMEs  of the STAR market so studied.}
        \label{fig:2}
\end{figure}

 We provide  figures  in order to visualize the data range for
$<$TIAX$>$2 and $<$TTA$>$2.
 on Fig.  \ref{Plot6TIAX2TTA2ranked}.
In these figures, the SMEs are ranked in increasing order of the
$y$-variable value. The range and statistical characteristics are
outlined in Table \ref{Table1}. Other displays, e.g., when the SMEs
are listed in alphabetical order, on the $x$-axis can be found in
Fig. \ref{fig:2}.

\begin{figure}[t]
            \includegraphics[width=0.6\textwidth, height=0.95\textwidth]
            {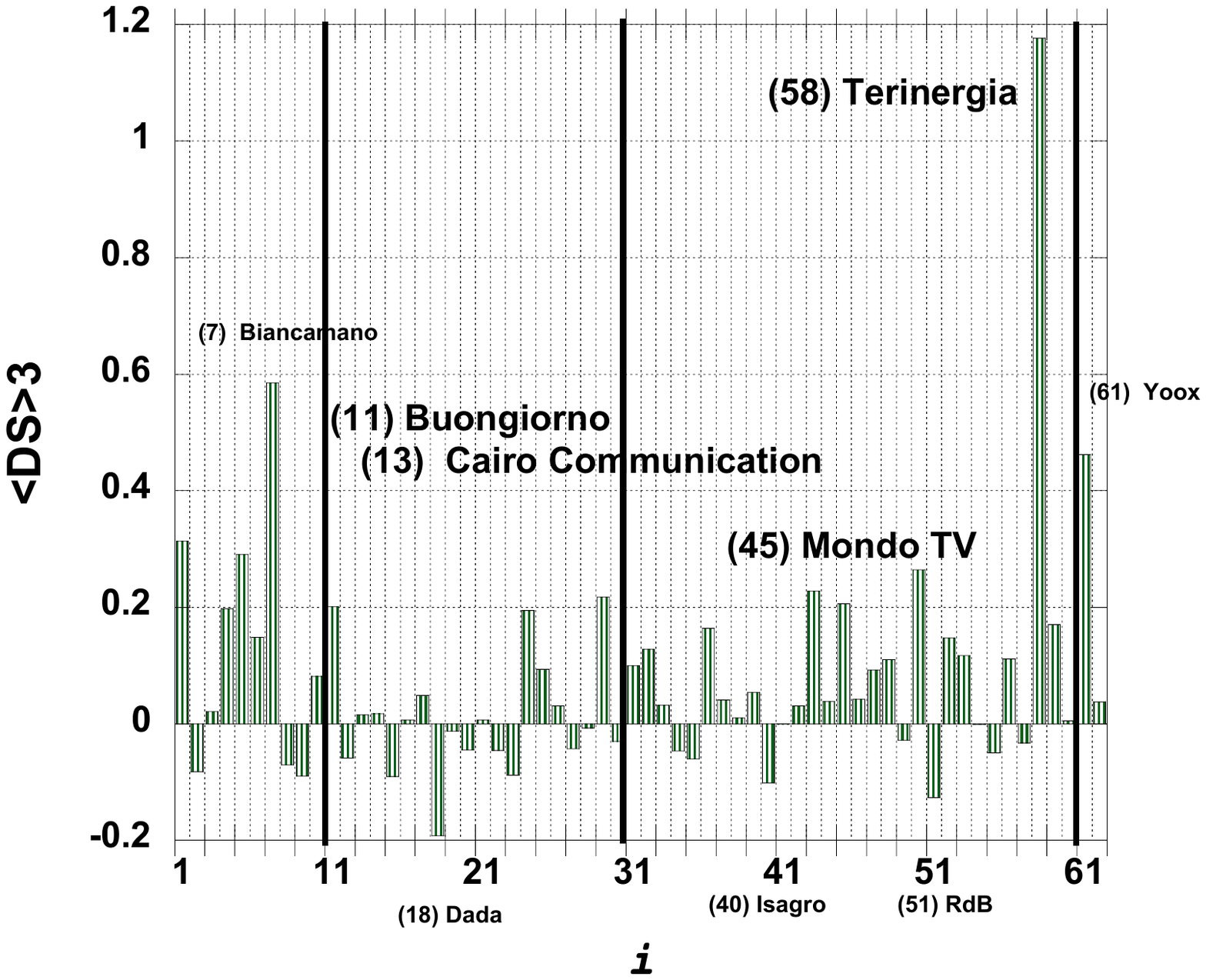}
           \label{fig:Plot6DS3}
               \includegraphics[width=0.6\textwidth, height=0.95\textwidth]
            {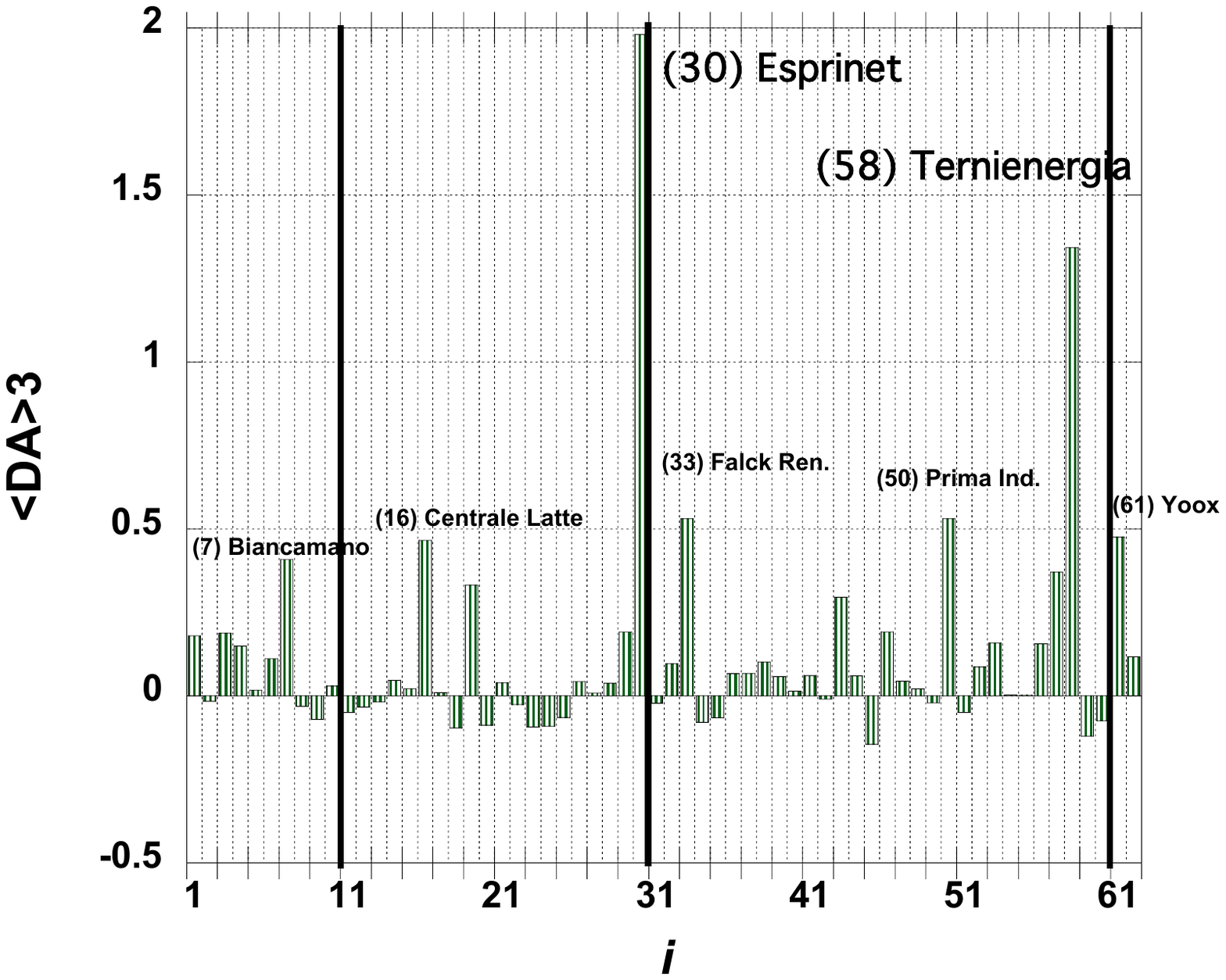}
            \label{fig:Plot5DA3}
         \caption{(color online)   Left panel:  sales variations  $<$DS$>$3.       Right panel: total assets variations  $<$DA$>$3;   thus each averaged over 3 years: [2008-2010],  - for the 62 SMEs , ranked in alphabetical order as in Table \ref{STARnames},
         particularly pointing to  a few relevant SMEs  of the STAR market so studied. }
        \label{fig:3}
        \end{figure}

\begin{figure}[t]
            \includegraphics[width=0.6\textwidth, height=0.95\textwidth]
            {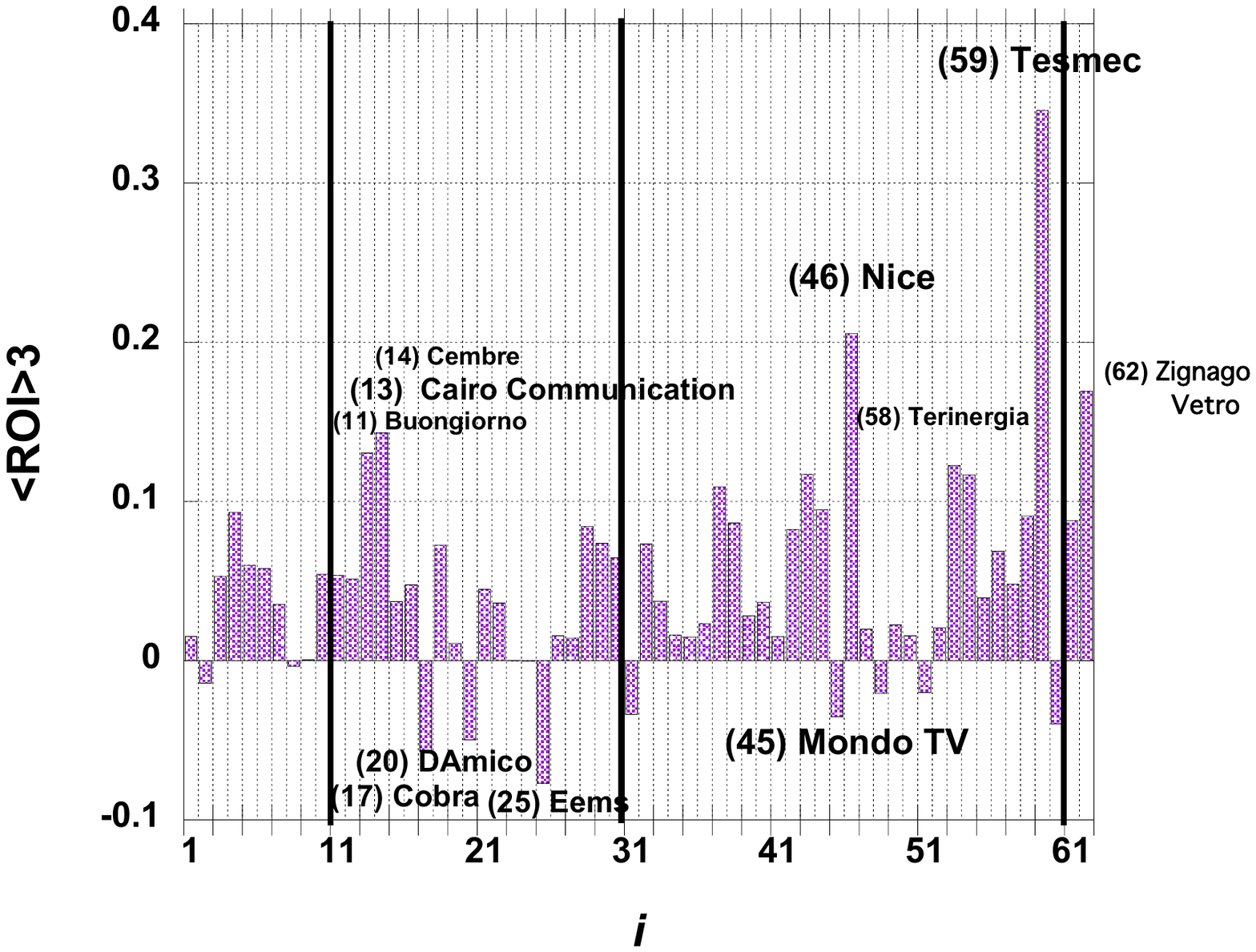}
           \label{fig:Plot3ROI3}
              \includegraphics[width=0.6\textwidth, height=0.955\textwidth]
            {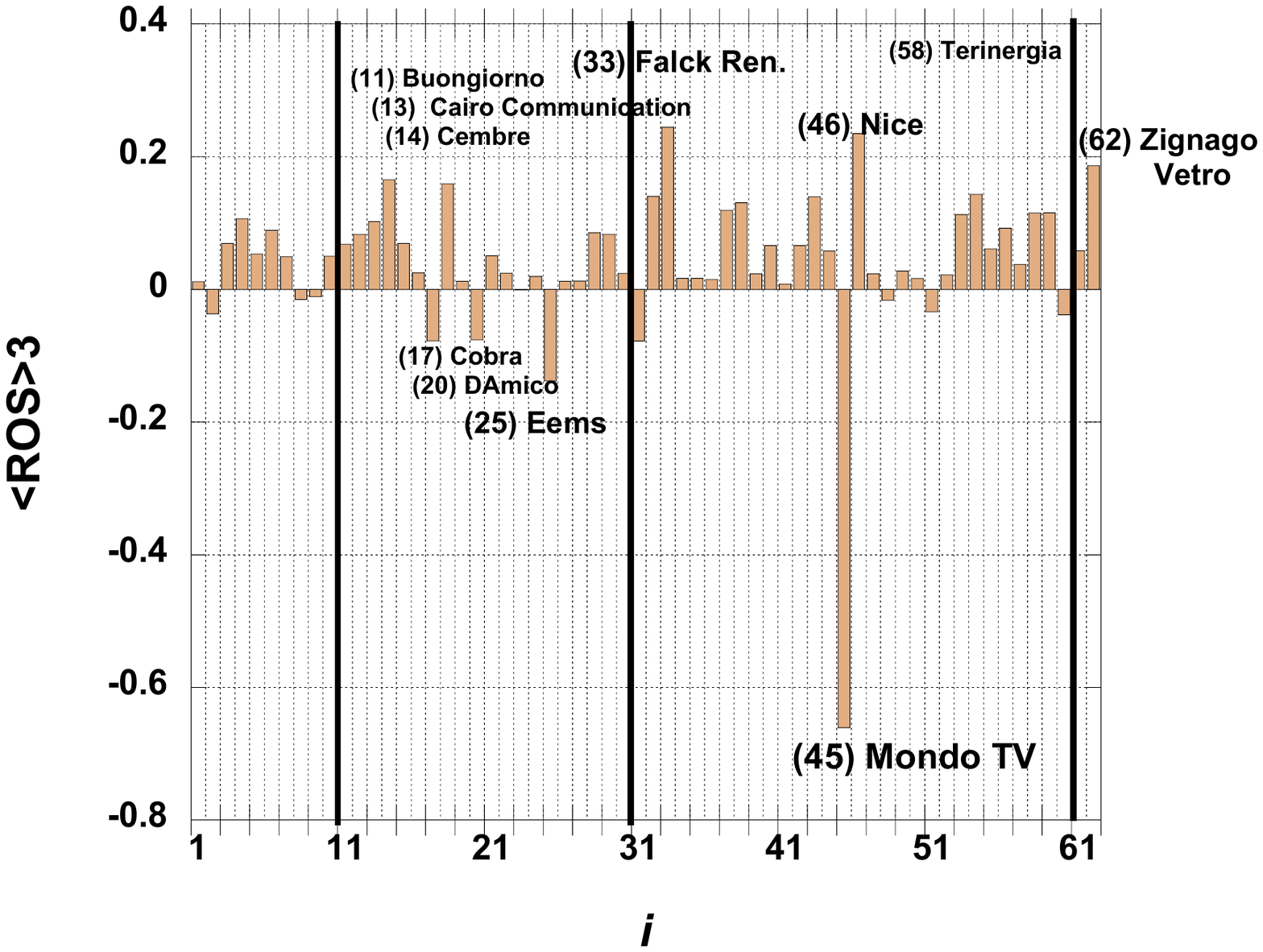}
            \label{fig:Plot2ROS3}
                    \caption{(color online)  Left panel: returns on investments  $<$ROI$>$3.     Right panel:  returns on sales  $<$ROS$>$3.  Thus each averaged over 3 years: [2008-2010],  - for the 62 SMEs , ranked in alphabetical order as in Table \ref{STARnames},
         particularly pointing to  a few relevant SMEs  of the STAR market so studied. }
        \label{fig:4}
\end{figure}

\begin{figure}[t]
             \includegraphics[width=0.6\textwidth, height=0.95\textwidth]
            {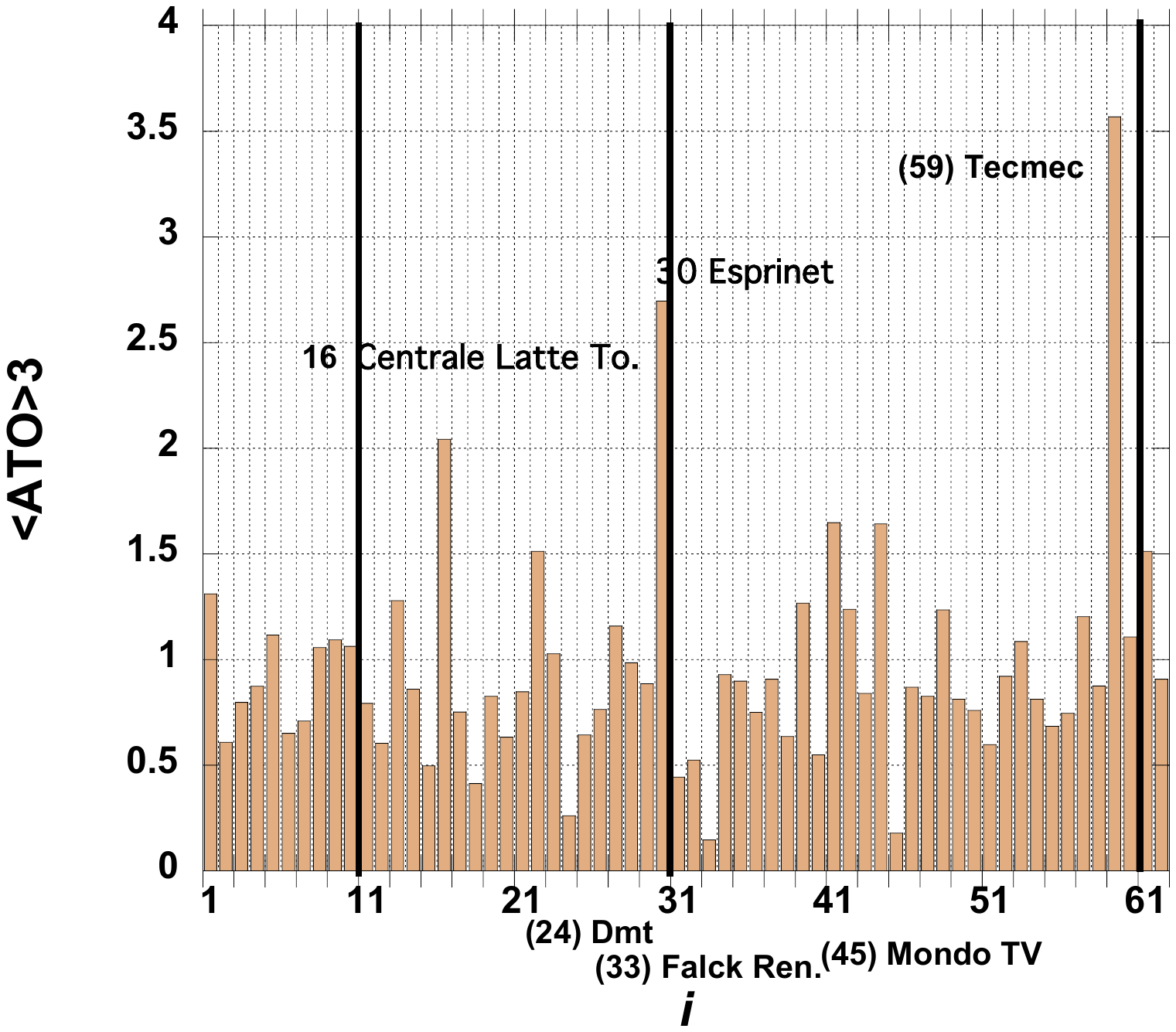}
           \label{fig:Plot7ATO3}
               \includegraphics[width=0.6\textwidth, height=0.95\textwidth]
            {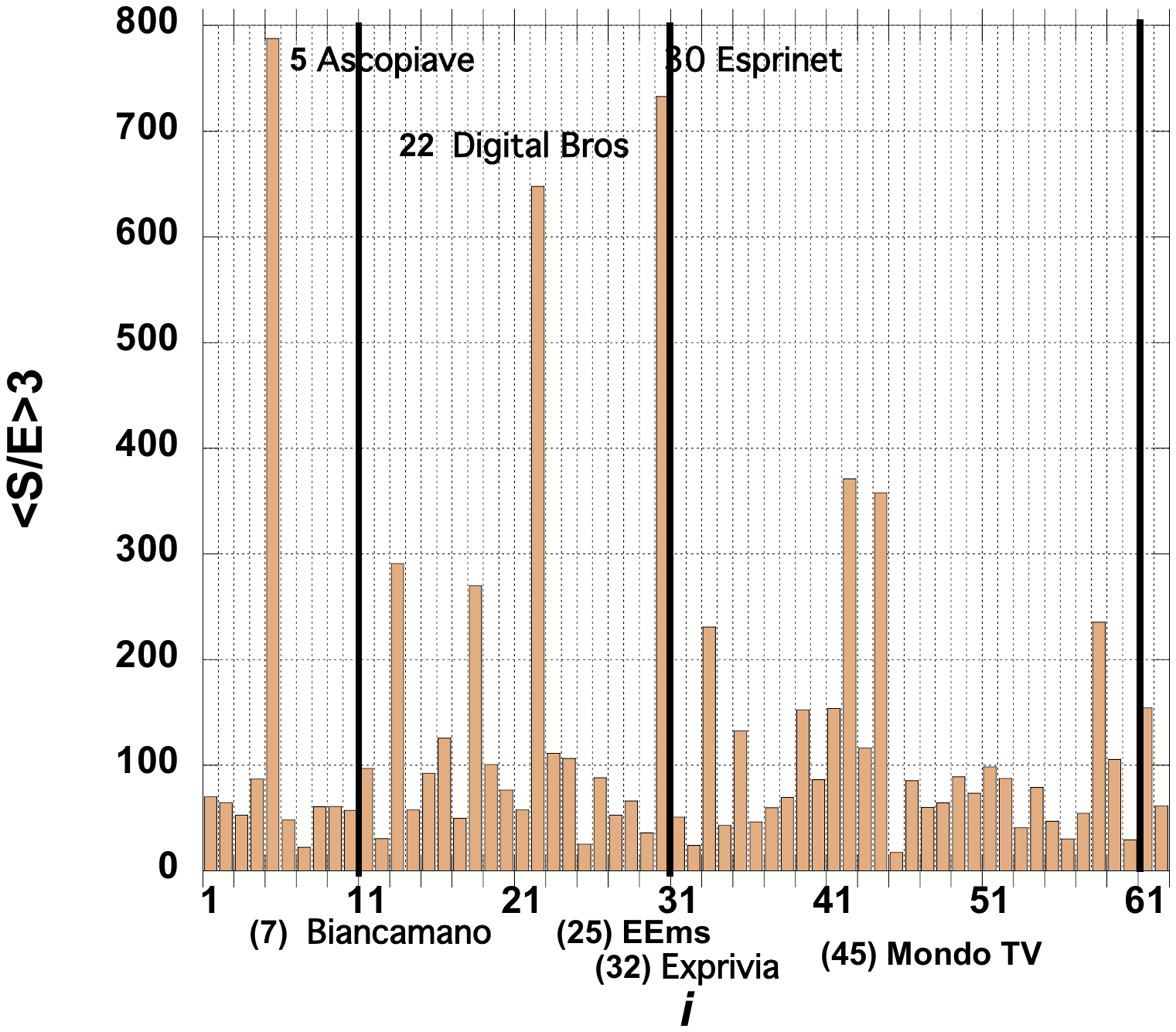}
            \label{fig:Plot8SpE3}
         \caption{(color online) Left panel:  asset turnovers $<$ATO$>$3. Right panel: sales per employee $<$S/E$>$3.  Thus each averaged over 3 years: [2008-2010],  - for the 62 SMEs , ranked in alphabetical order as in Table 1\ref{STARnames}, 
         particularly pointing to  a few relevant SMEs  of the STAR market so studied}
        \label{fig:5}
\end{figure}

\begin{figure}[t]
       % \centering
      %   \subfigure{    \centering
               \includegraphics[width=0.6\textwidth, height=0.95\textwidth]
               {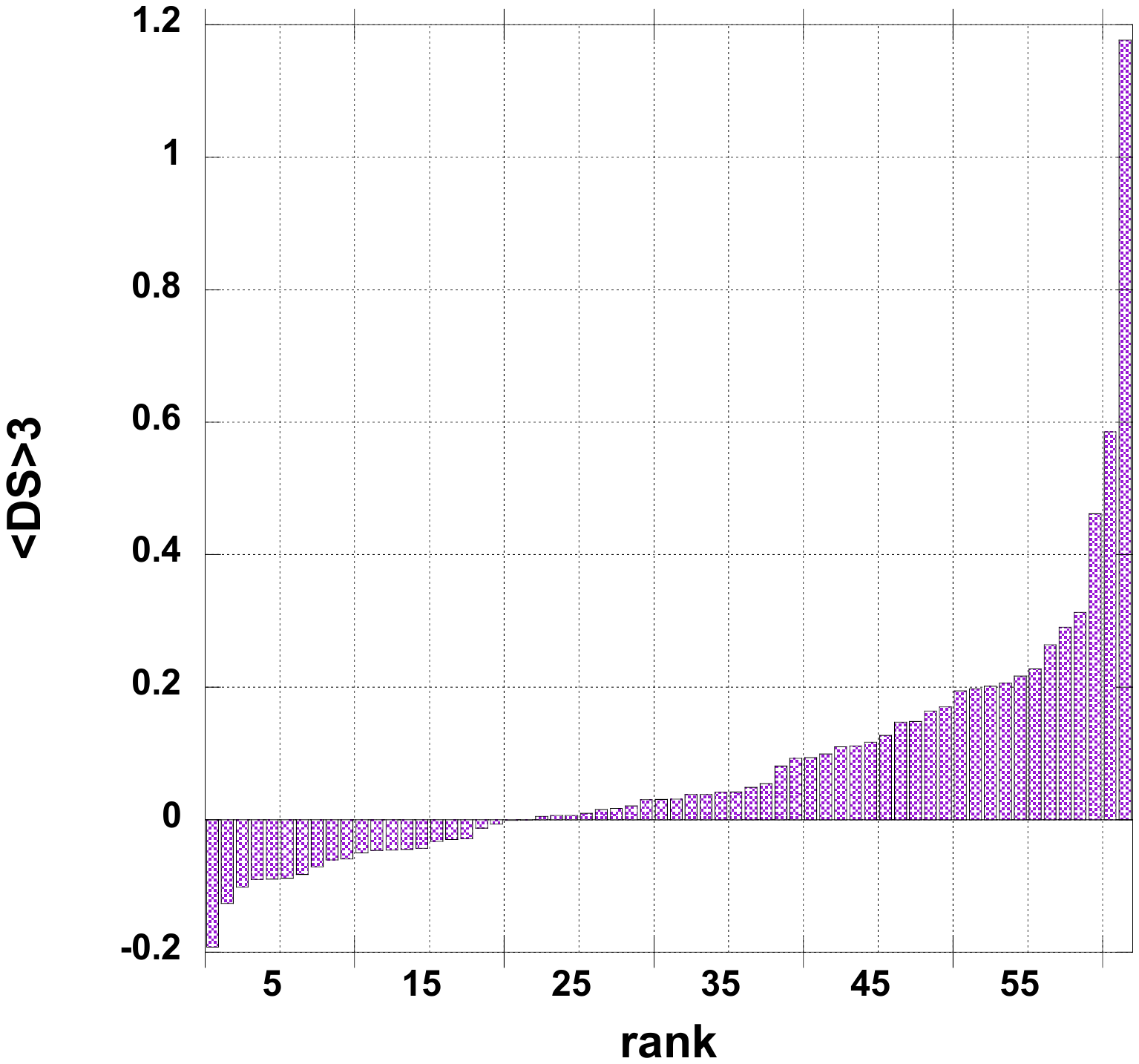}
           \label{fig:Plot6DS3ranked}
      %  }
        ~ %add desired spacing between images, e. g. ~, \quad, \qquad, \hfill etc.
   %     \subfigure{
            \includegraphics[width=0.6\textwidth, height=0.95\textwidth]
                     {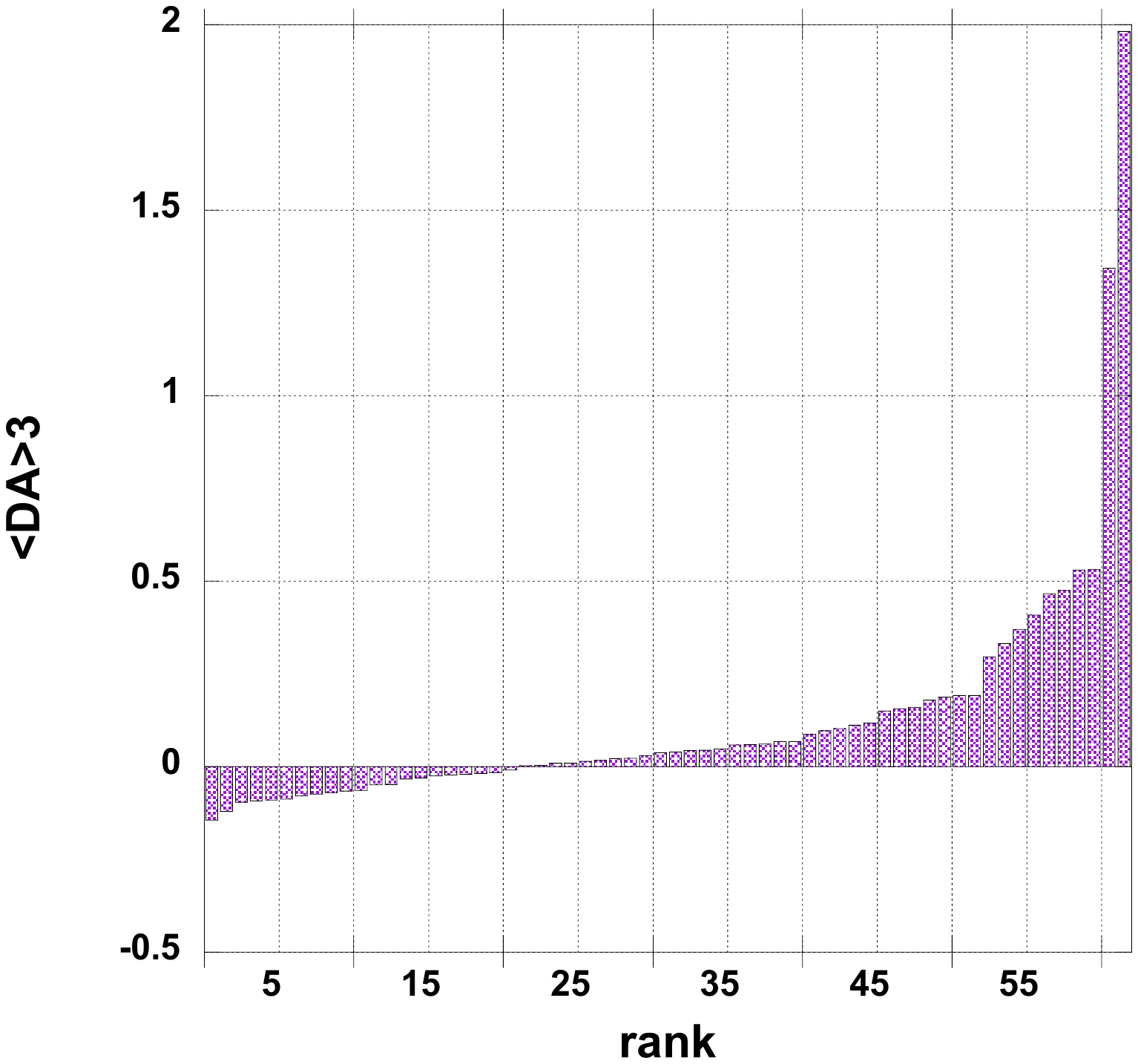}
            \label{fig:Plot6DA3ranked}
      %  }
         \caption{(color online)   Left panel:  sales variations $<$DS$>$3.     Right panel:  total assets variations  $<$DA$>$3;  both ranked in increasing order,  - for the 62 SMEs,    particularly pointing to  a few relevant SMEs  of the STAR market so studied.}
        \label{fig:6}
\end{figure}

\begin{figure}[t]
               \includegraphics[width=0.6\textwidth, height=0.955\textwidth]
            {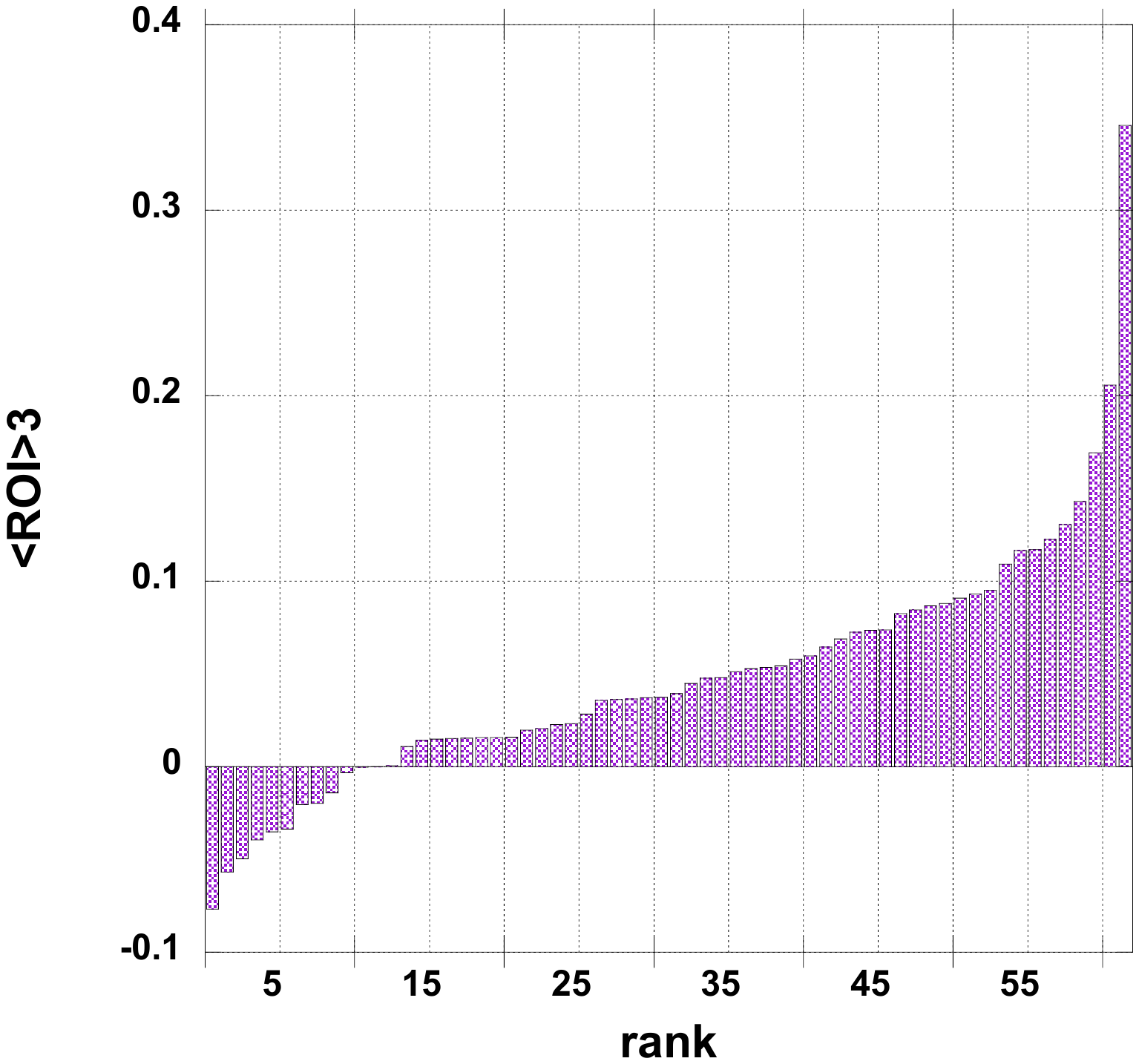}
          %  \label{fig:Plot6ROI3ranked}
     ~          \includegraphics[width=0.6\textwidth, height=0.95\textwidth]
            {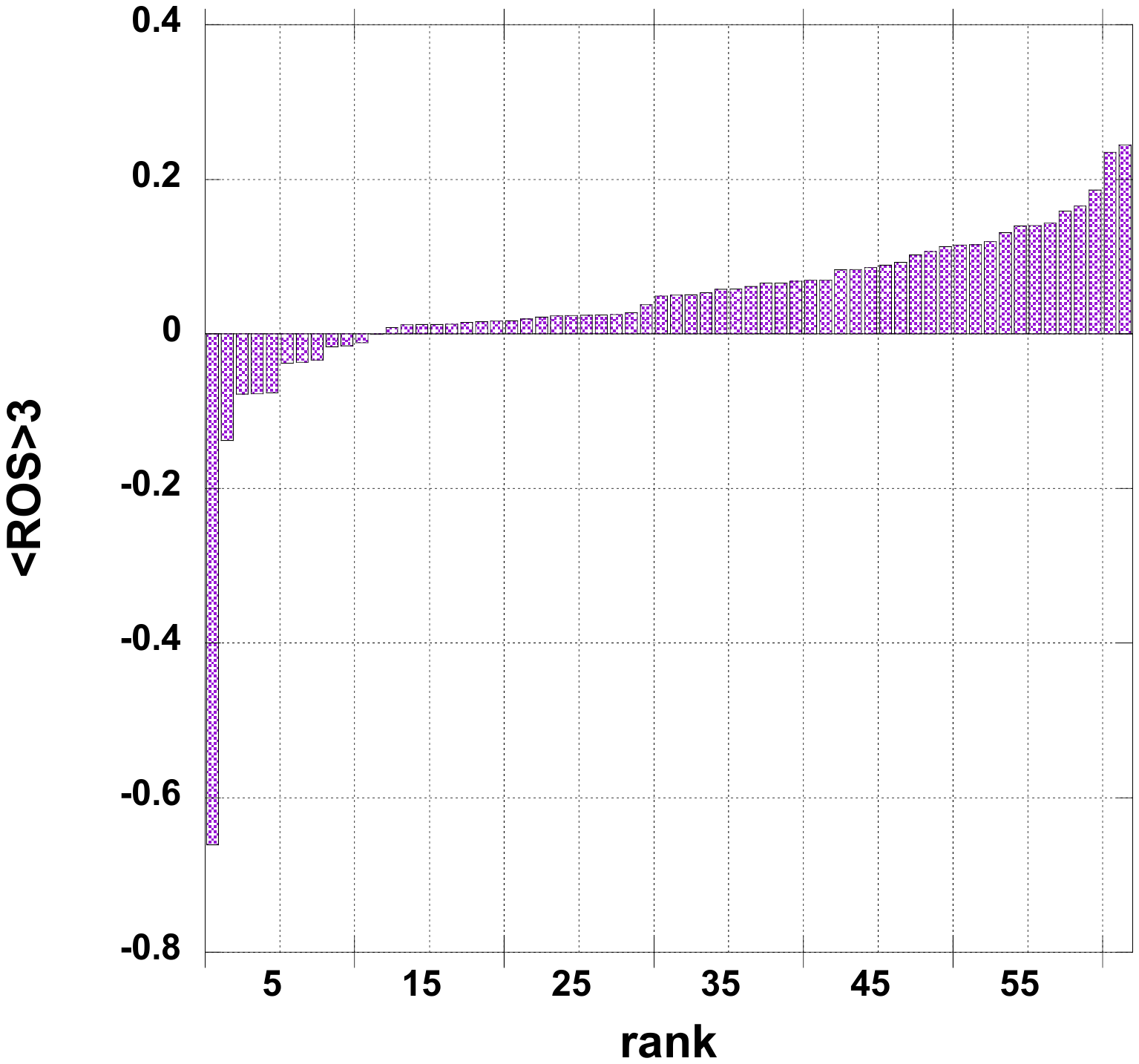}
         \caption{(color online)  Left panel: returns on investments   $<$ROI$>$3.    Right panel:  returns on sales $<$ROS$>$3;  both ranked in increasing order,  - for the 62 SMEs,    particularly pointing to  a few relevant SMEs  of the STAR market so studied.}
        \label{fig:7}
\end{figure}

\begin{figure}[t]
  \includegraphics[width=0.6\textwidth, height=0.95\textwidth]
  {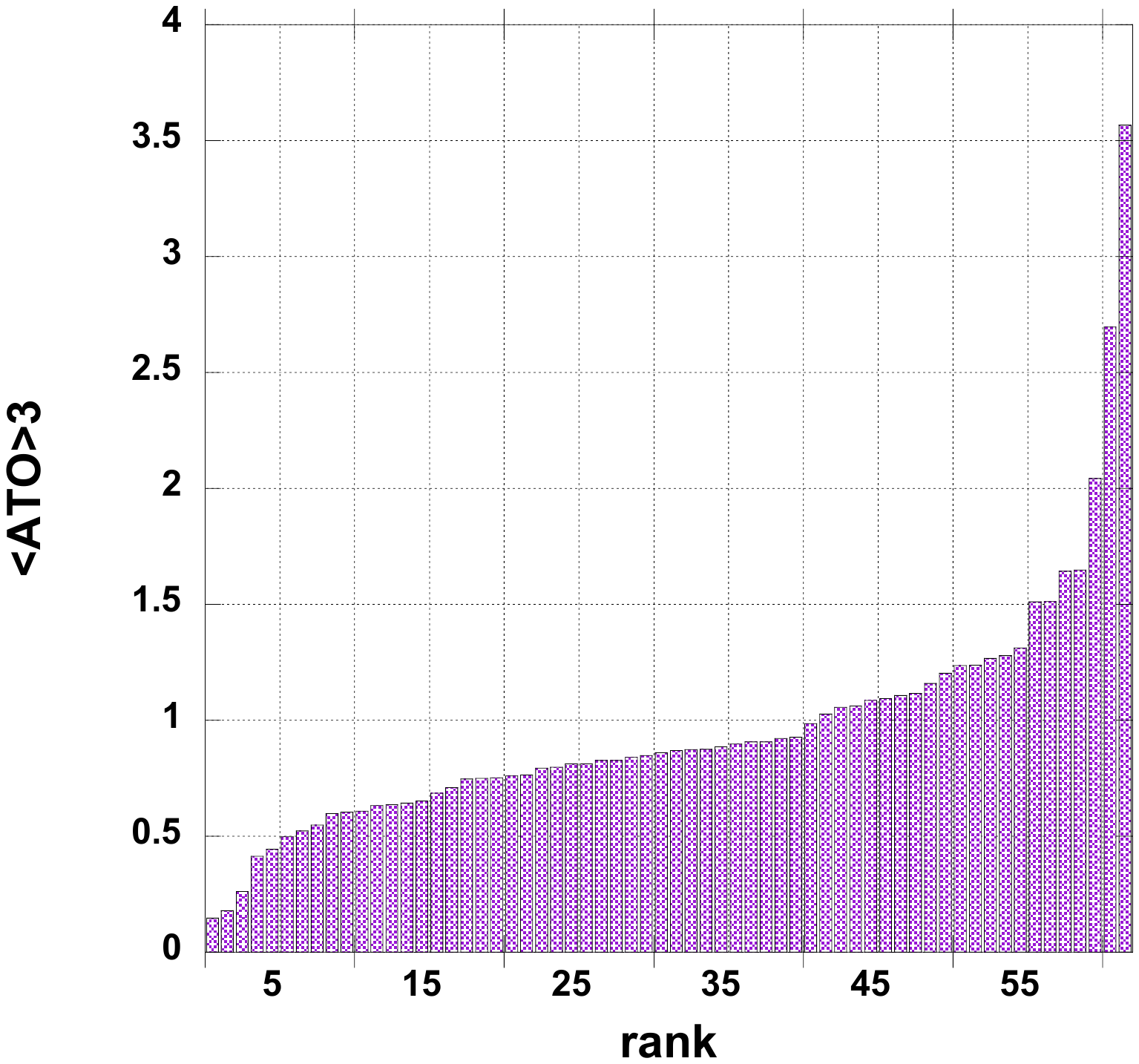}
           \label{fig:Plot6ATO3ranked}
 ~
            \includegraphics[width=0.6\textwidth, height=0.95\textwidth]
                        {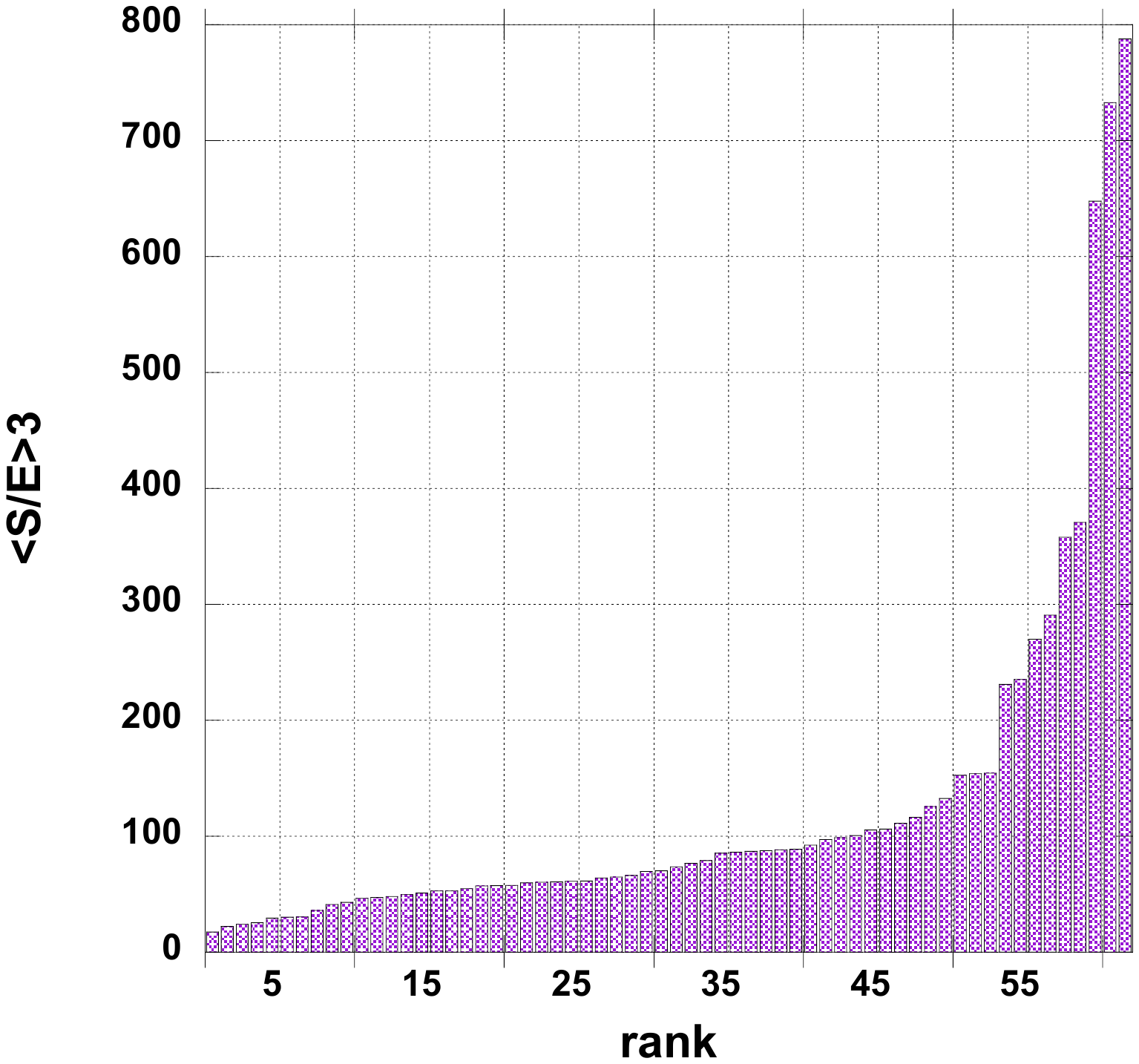}
            \label{fig:Plot6SpE3ranked}
         \caption{(color online)  Left panel: asset turnovers $<$ATO$>$3.     Right panel: sales per employee $<$S/E$>$3;  both ranked in increasing order,  - for the 62 SMEs,    particularly pointing to  a few relevant SMEs  of the STAR market so studied. }
        \label{fig:8}
\end{figure}

\clearpage
 Next, let us display the performance variables  \underline{averaged over 3 years}, [2008-2010]:
\begin{itemize}
\item $<$DS$>$3  for the sales variations,
\item $<$DA$>$3 for  the total assets variations,
%\item  $<$DL$>$3 represents the average of the number of employees variations over 3 years: [2008-2010];
\item $<$ROI$>$3 for ROI, %over 3 years: [2008-2010]; %Fig. \ref{Plot6ROI3}.
\item $<$ROS$>$3 for ROS, %over 3 years: [2008-2010];
\item $<$ATO$>$3 for the asset turnovers, % over 3 years: [2008-2010];
\item $<$S/E$>$3 for  the   sales per employee, % over such 3 years
\end{itemize}
either when companies are listed in alphabetical order, as in Figs. \ref{fig:3}-\ref{fig:5},
or ranked  in increasing order of  the relevant variable, as in Figs. \ref{fig:6}-\ref{fig:8}.
%Statistical characteristics are given in Table \ref{Table1}.

Statistical characteristics for the distributions of the  averaged
innovation and  performance indicators are found in Table
\ref{TablestatSKSK}.

\clearpage

        \begin{figure}[t]
  \includegraphics[width=0.6\textwidth, height=0.85\textwidth]
  {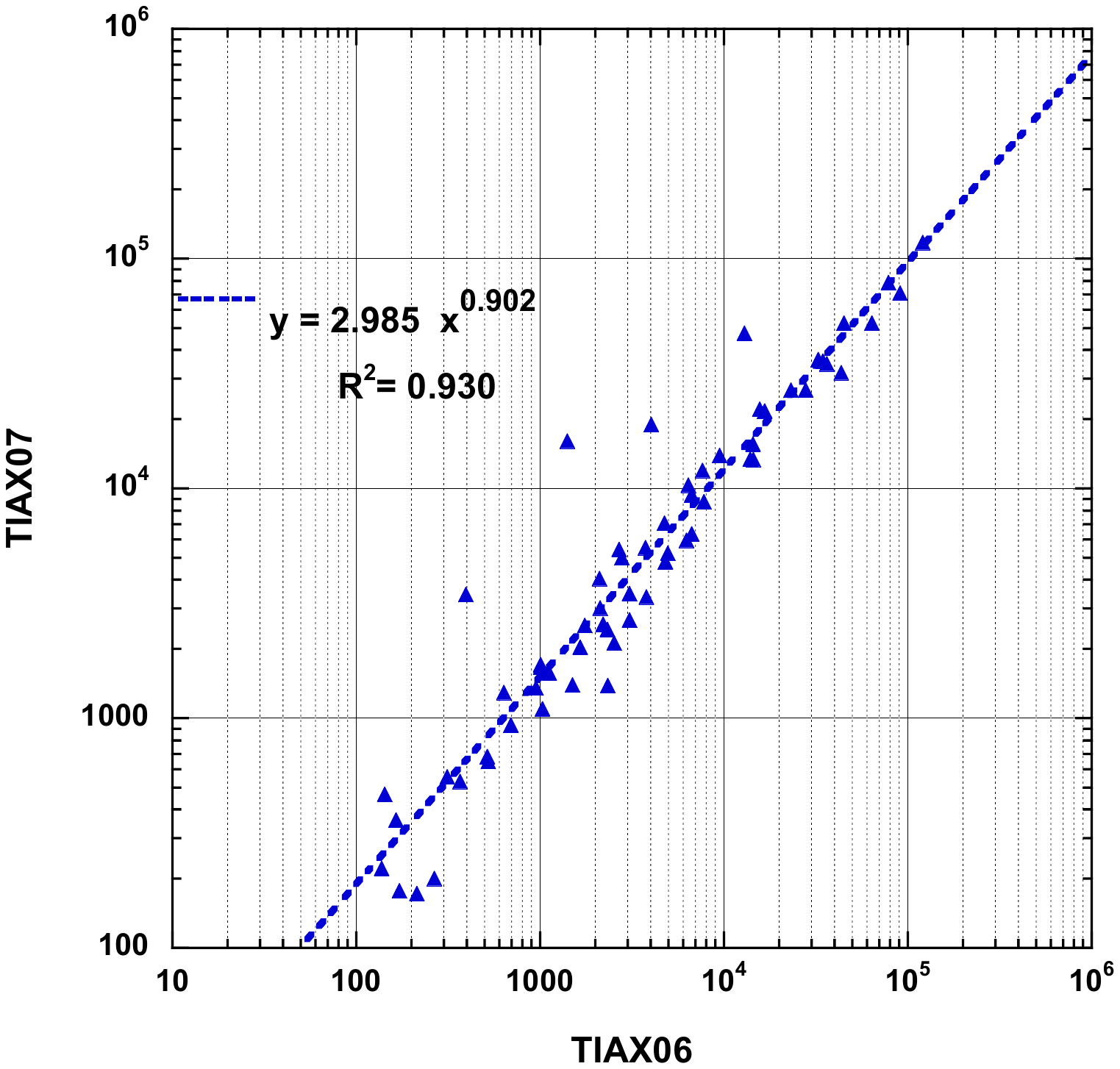}
           \label{fig:Plot11TIAX06TIAX07lolo}
 %~
            \includegraphics[width=0.6\textwidth, height=0.85\textwidth]
                        {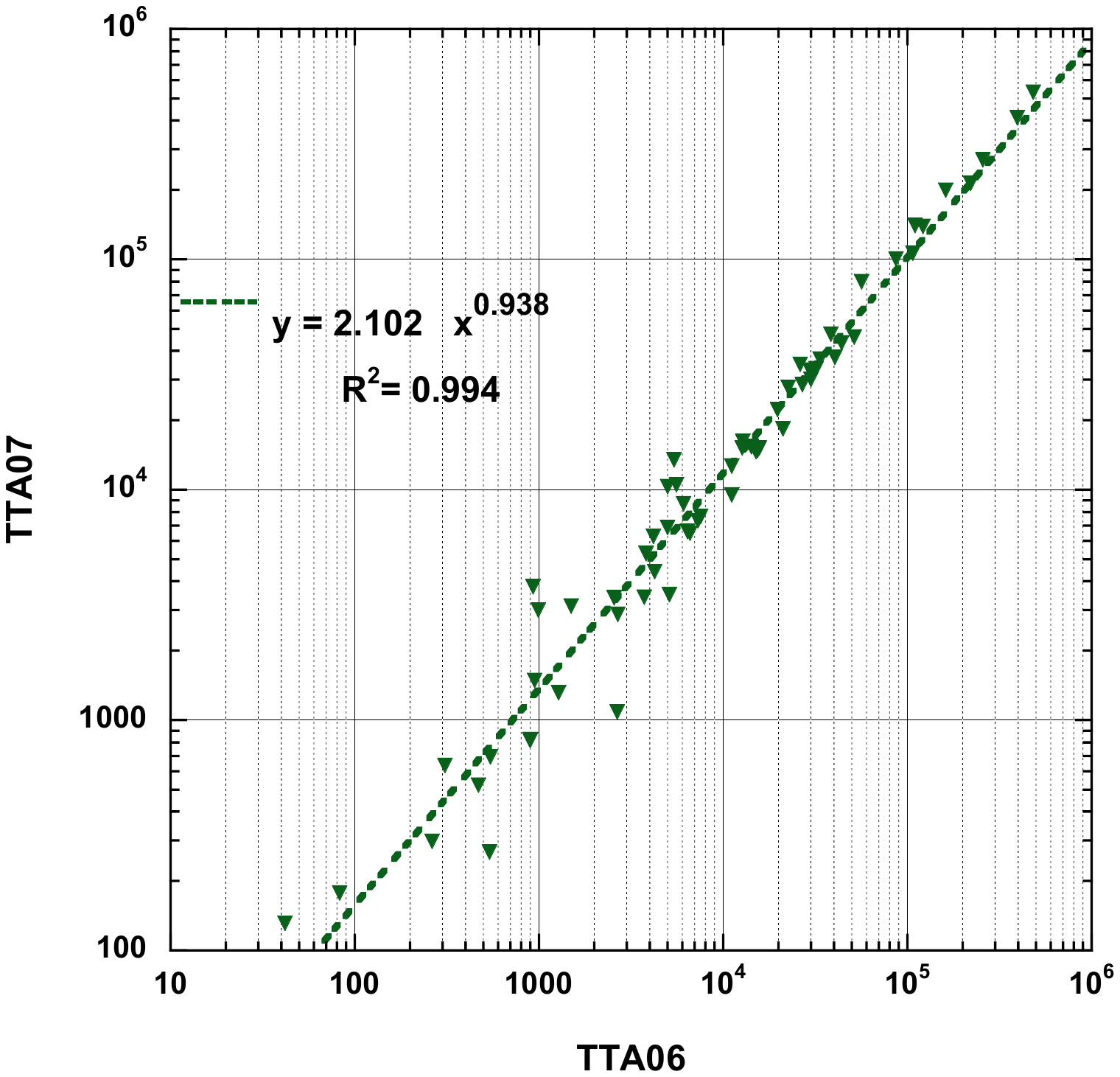}
            \label{fig:Plot10TTA06TTA07lolo}
         \caption{Power law regression analysis for (color online) left panel: TIAX07 vs. TIAX06 and right panel: TTA07 vs. TTA06,   for the 62 SMEs. }
        \label{fig:9}
\end{figure}

\section{Discussion}\label{Discussion}
\begin{figure}[t]
  \includegraphics[width=0.6\textwidth, height=0.85\textwidth]
  {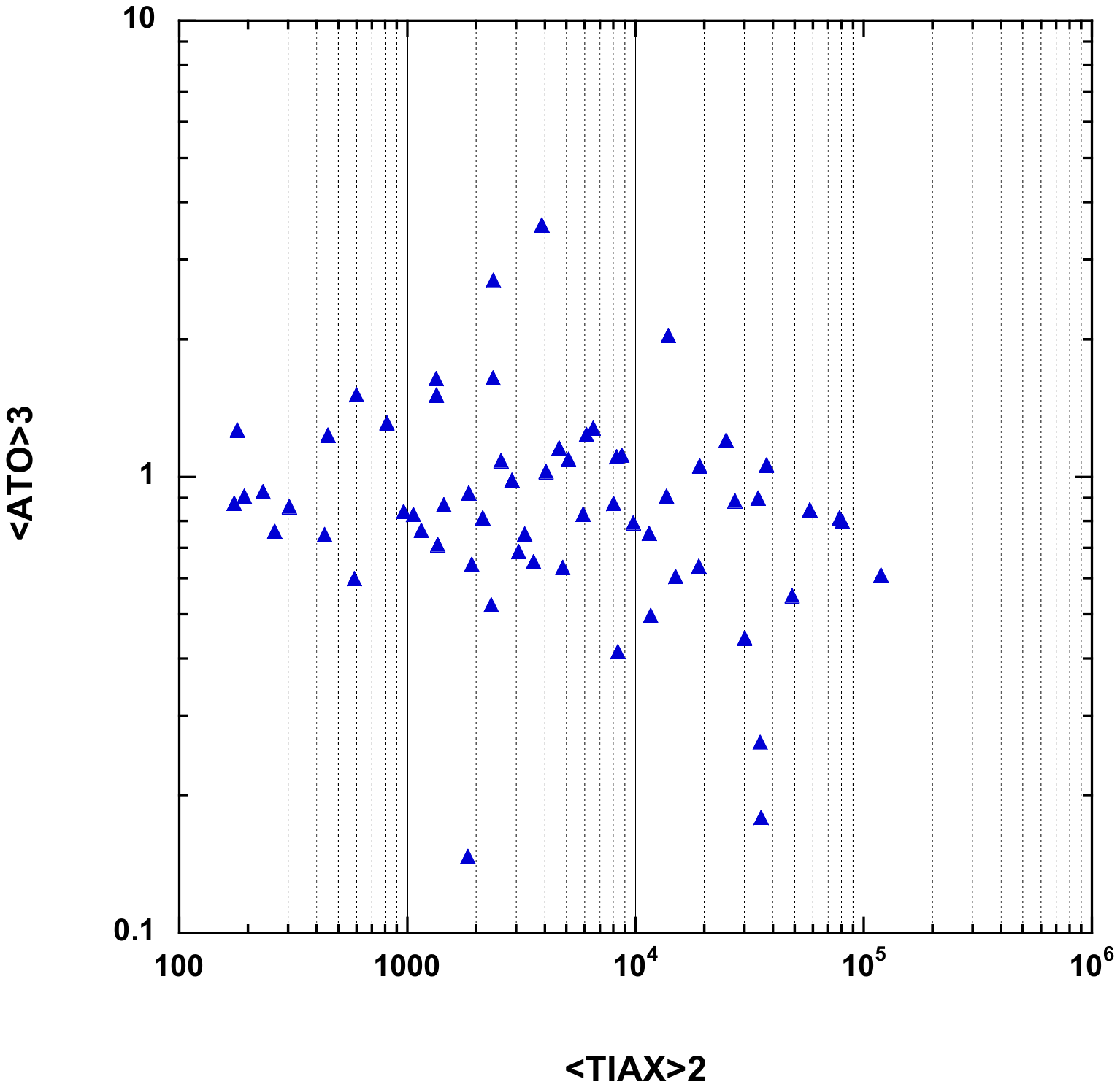}
           \label{fig:Plot15ATO3TIAX2lolo}
 %~
            \includegraphics[width=0.6\textwidth, height=0.85\textwidth]
                        {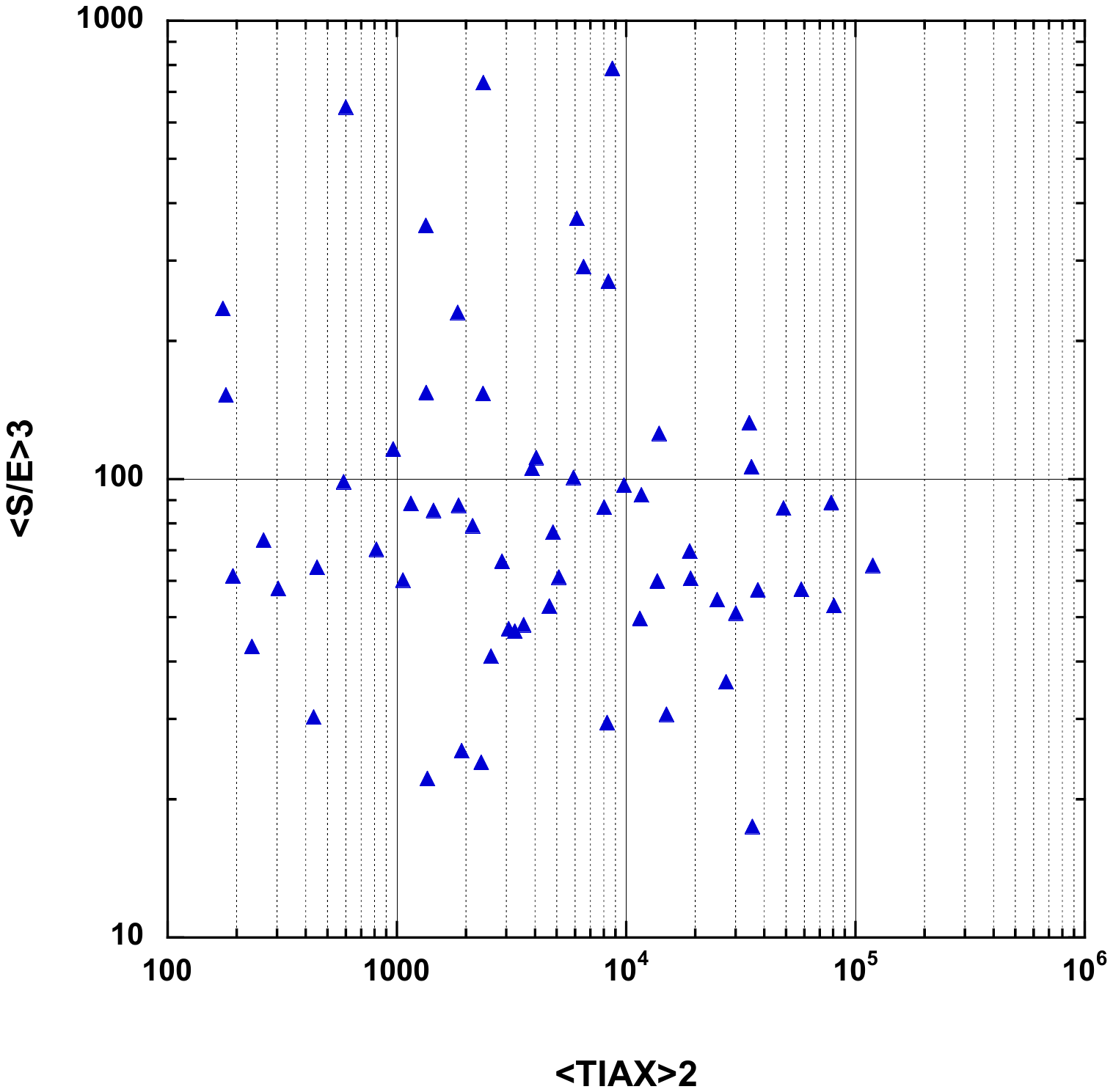}
            \label{fig:Plot16SpE3TIAX2lolo}
         \caption{Searching for correlations: (color online) left panel: $<$ATO$>$3 vs. $<$TIAX$>$2; right panel:  $<$S/E$>$3 vs. $<$TIAX$>$2.   }
        \label{fig:10}
\end{figure}

\begin{figure}[t]
  \includegraphics[width=0.6\textwidth, height=0.85\textwidth]
  {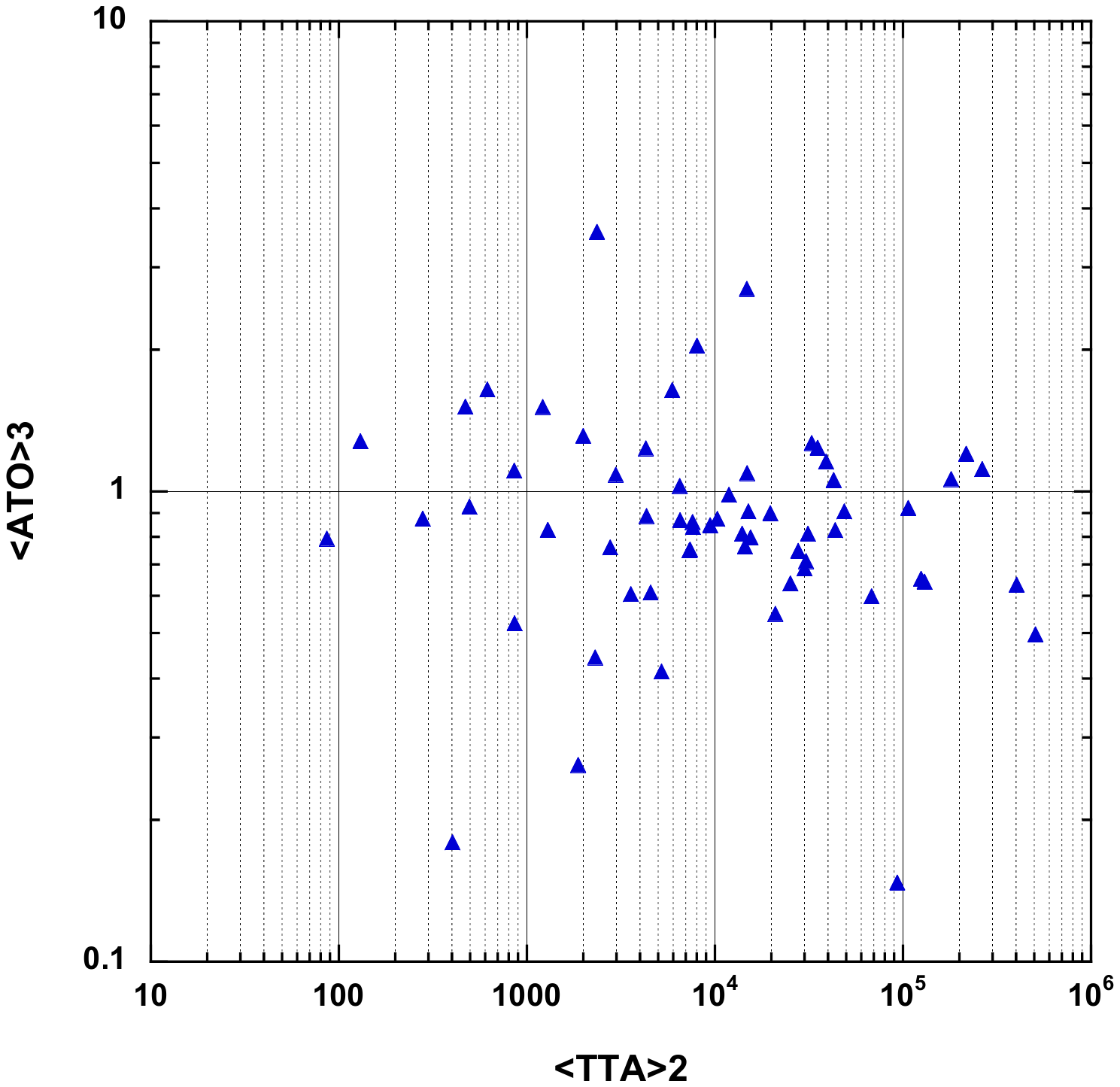}
           \label{fig:Plot13ATO3TTA2lolo}
 %~
            \includegraphics[width=0.6\textwidth, height=0.85\textwidth]
                        {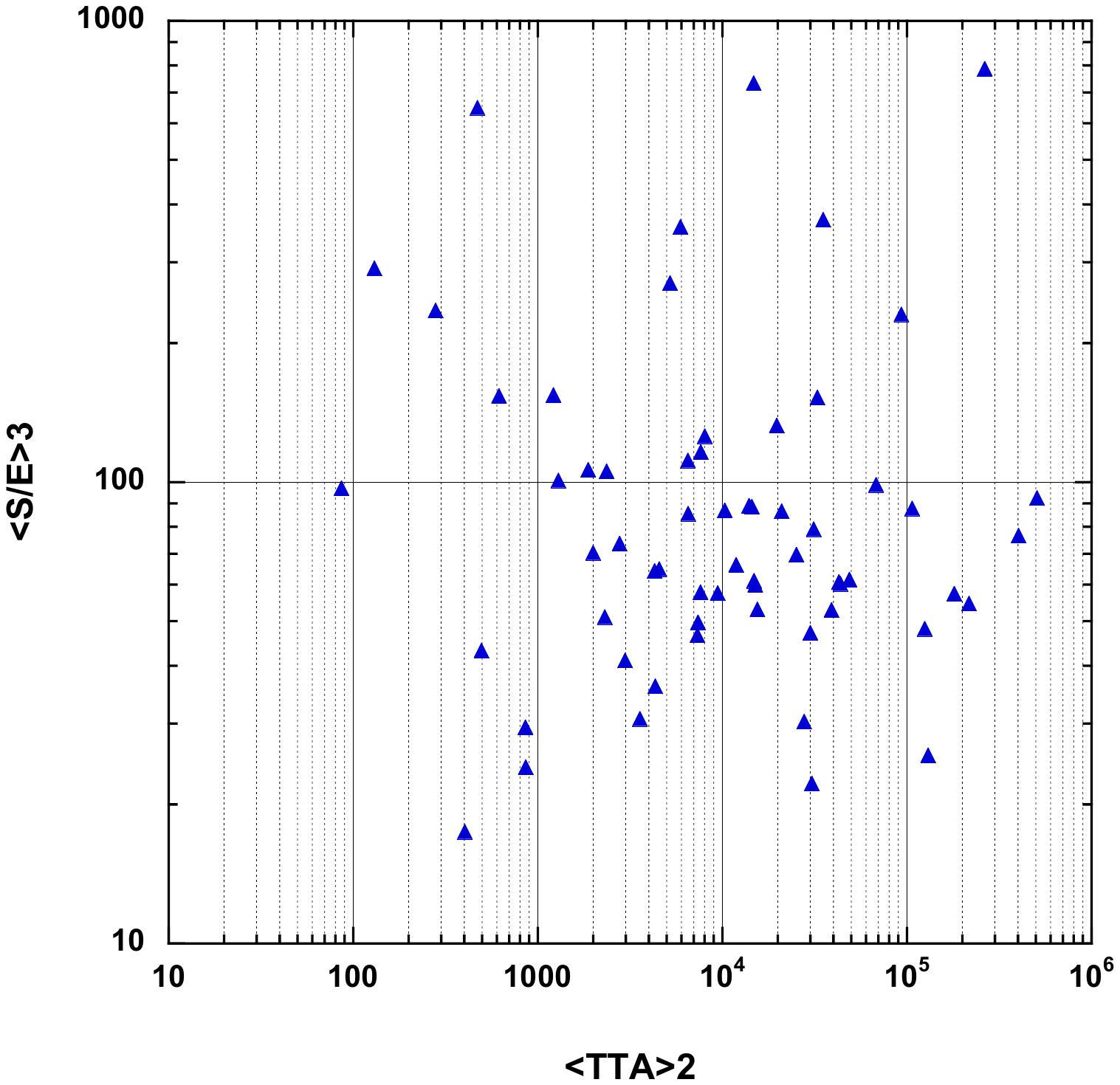}
            \label{fig:Plot14SpE3TTA2lolo}
         \caption{Searching for correlations: (color online) left panel: $<$ATO$>$3 vs. $<$TTA$>$2; right panel: $<$S/E$>$3 vs. $<$TTA$>$2.   }
        \label{fig:11}
\end{figure}

Many correlations can be searched for; beside  those\footnote{Notice
that the relationships are  not exactly linear.} between   TTA06 and
TTA07, or TIAX06 and TIAX07, shown on Fig. \ref{fig:9}, one may
consider
those between the averaged variables, like % Fig. 2 interesting ones  are  for
\begin{itemize}
%\item $<$ROI$>$3   vs.  $<$ROS$>$3
%\item $<$DA$>$3   vs.  $<$DS$>$3
\item $<$DS$>$3  vs. $<$TTA$>$2 % Fig. 4 v
\item $<$DA$>$3 vs. $<$TTA$>$2 % Fig  3 v
\item $<$ROI$>$3 vs. $<$TTA$>$2 % Fig. 5 v
\item $<$ROS$>$3 vs. $<$TTA$>$2 % Fig.6 v
\end{itemize}
which can be read in Figs. 8-11, in Ausloos et al. (2018b), whence are not reproduced here. Nevertheless, for completeness, we show
\begin{itemize}
\item $<$ATO$>$3 vs. $<$TIAX$>$2 % Fig. 5
\item $<$S/E$>$3 vs. $<$TIAX$>$2 % Fig.6
\item $<$ATO$>$3 vs. $<$TTA$>$2 % Fig. 5
\item $<$S/E$>$3 vs. $<$TTA$>$2 % Fig.6
\end{itemize}
on Fig. \ref{fig:10} and Fig. \ref{fig:11}.

It should be apparent that the data  looks pretty scattered, suggesting a "more sophisticated" approach for reaching some
conclusion. As an intermediary remark, observe that $<$ATO$>$3  and
$<$S/E$>$3  are all positive; this is not the case for
 $<$DS$>$3,  $<$DA$>$3,  $<$ROI$>$3,  and  $<$ROS$>$3;  some SMEs have negative values in the latter cases; see Fig. \ref{fig:6} and on 
 Fig. \ref{fig:7}, for examples.

\clearpage

\subsection{A brief description of the Voronoi tessellation}
\label{briefvoronoi}

The Voronoi tessellation is a method for decomposing a metric space in non-overlapping subsets. Such a methodology dates back to Ren\' e Descartes, who informally described it in his 
Principia Philosophiae (Descartes, 1644). Later, it was formalized in the context of the multidimensional real spaces (Voronoi, 1908).
The principles behind the conceptualization of the Voronoi tessellation are grounded on the criterion used for decomposing the space. Some specific points -- the so-called "centroids" or "seeds" -- are initially selected. In our context, we refer to a finite number of centroids. Then, 
the space is partitioned into regions, according to the distances from the seeds. Specifically, each point of the space is assigned to the peculiar centroid which is closer to it. In so doing, the points assigned to a given centroid form a region which contains the centroid itself and does not overlap with other regions/centroids. 
When all the points of the space are assigned to a specific centroid, then the space appears visually as "tesselled"; this intuitively suggests why one refers to the "Voronoi tessellation". The distance employed for the tessellation procedure can be selected in a number of ways, and it is based on the metric. Here and in most   applications , -- and also in the original Voronoi's paper,   the considered metric space is the multidimensional Euclidean space. Thus, the natural Voronoi distance is the Euclidean one.

In the  present  application, we refer to bidimensional Euclidean spaces;  the coordinates of the considered points and centroids are $x$ and $y$ variables.

 \subsection{Voronoi correlations approach}\label{correlations}
\begin{figure}   %12
 %   \begin{center}%centering
   \includegraphics[width=0.55\textwidth, height=0.5\textwidth]{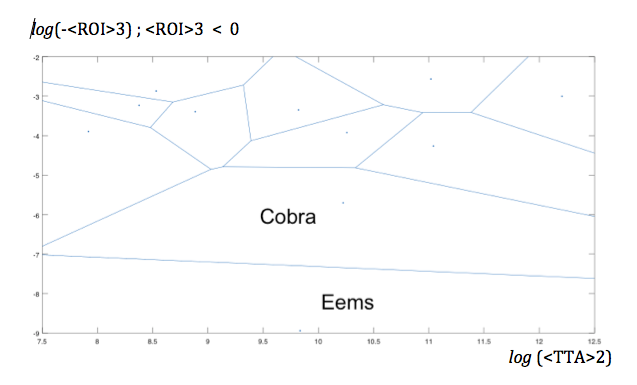} %{docum12alogTTA2logROI3-.pdf}
   %  ~
            \includegraphics[width=0.55\textwidth, height=0.5\textwidth] {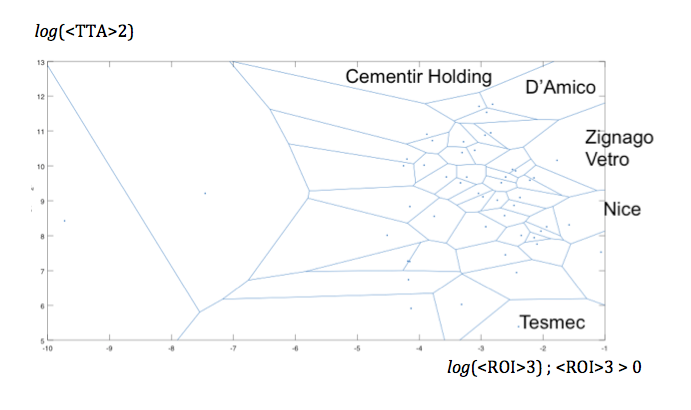} %            {docum12blogROI3+logTTA2.pdf}
\caption{  Voronoi    tessellation of   [log($<$ROI$>$3), log($<$TTA$>$2)] plane. Left panel:   when  $<$ROI$>$3  $<$ 0.  Right panel: when $<$ROI$>$3  $>$ 0 for the 62 SMEs
 discussed in the text. A few specific SMEs are pointed out. }
\label{fig:VoronoiROI3-+}
%\end{center}
\end{figure}

\begin{figure}   %13
%    \begin{center}%centering
\includegraphics[width=0.55\textwidth, height=0.5\textwidth]{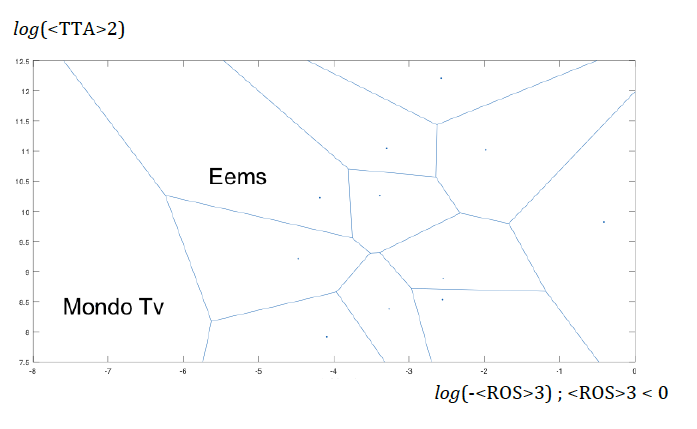} % {docum13alogROS3-logTTA2}
   %  ~
      \includegraphics[width=0.55\textwidth, height=0.5\textwidth] {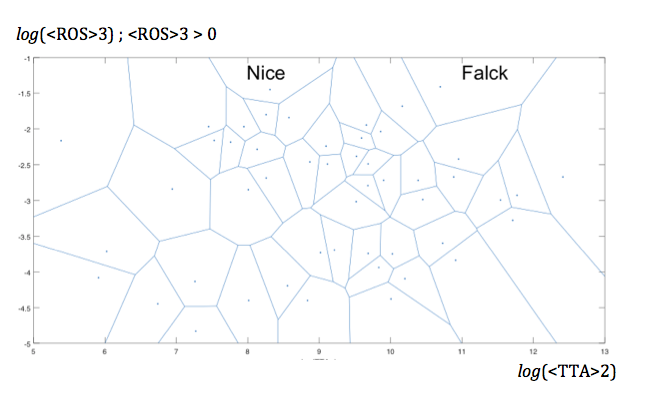} %            {docum13blogTTA2logROS3+.pdf}
\caption{  Voronoi    tessellation of   [log($<$ROS$>$3), log($<$TTA$>$2)] plane. Left panel:  when $<$ROS$>$3 $< 0$; Right panel: when $<$ROS$>$3  $> 0$ for the 62 SMEs discussed
 in the text. A few specific SMEs are pointed out.}
\label{fig:VoronoiROS3-+}
%\end{center}
\end{figure}

In the context of Voronoi tessellation of the bidimensional Euclidean space, the $x$ and $y$ axes correspond thereafter to one of the innovation
($\mathcal{I}$) and one of  the performance  ($\mathcal{P}$)
variables respectively. It is easily understood that counting only
correlations between   ($\mathcal{I}$) and performance
($\mathcal{P}$) variables, one has 12 displays; the more so if one
considers the log of the variables for display readability
("scaling"),  since  as pointed out the absolute value of several
($\mathcal{P}$) variables has to be taken before log-scaling,
leading to  20 Voronoi maps. This seems to be fine for completeness,
but too much for illustrating the  purpose and its pedagogical
approach at this time. Thus, only a few cases are illustrated
thereafter: Figs. \ref{fig:VoronoiROI3-+} -\ref{fig:VoronoiROS3-+},
for  the relationship between  $<$TTA$>$2 and  $<$ROI$>$3  or
$<$ROS$>$3. For readability, the $x$ and $y$ axes differ (are
flipped)   depending on  the figure panel. However, this allows to
observe the size of the extreme regions, in which, in some sense,
the whole market is divided.

%\clearpage

\subsection{Voronoi clustering approach }\label{Voronoi}

In the Voronoi clustering approach, for avoiding scale effect,  the
variables of interest  are first normalized.  For each company $j=1,
\dots, 62$, we define:
\begin{equation}
\label{xj} \bar{x}_j=\frac{x_j-m_x}{M_x-m_x},
\end{equation}
where   $x_j$ represents
the value of the variable $x$ for the $j$-th company and
$$
m_x=\min\limits_{j=1, \dots, 62} x_j,\qquad M_x=\max\limits_{j=1,
\dots, 62} x_j.
$$

Next, in search of clusters, the centroids of the Voronoi tessellation are  {\it a priori} defined by positive numbers, $\{\phi_h\}_{h=1}^H$ and $\{\psi_k\}_{k=1}^K$,
where $H$ and $K$ are {\it a priori} chosen integers. %, for the case of innovation and performance variables, respectively.
%The ranges of variation of the centroids depends on the selected distance measure, as we will see soon.
%Here below $H=K=4$.

Moreover, we introduce  a  \underline{weighted Euclidean distance}, for  each innovation variable ($\mathcal{I}$),   through

\begin{equation}
\label{dalpha} d_{\mathcal{I}}(j,\phi_h)=\sum_{x \in \mathcal{I}}
\alpha_x (\bar{x}_j-\phi_h)^2,
\end{equation}
for each centroid $\phi_h$ and where the $\alpha$'s  are the
nonnegative weights of the norm, which can differ depending on the centroid, but so that
$$
\sum_{x \in \mathcal{I}} \alpha_x =1.
$$
Analogously, for  each performance variable  ($\mathcal{P}$), we define:
\begin{equation}
\label{dbeta} d_{\mathcal{P}}(j,\psi_k)=\sum_{x \in \mathcal{P}}
\beta_x (\bar{x}_j-\psi_h)^2,
\end{equation}
for each centroid $\psi_k$, imposing
$$
\sum_{x \in \mathcal{P}} \beta_x =1.
$$
In so doing, all  distances are $0 \leq d_{\mathcal{I}}(j,\phi_h),
d_{\mathcal{P}}(j,\psi_k) \leq 1$, for each company $j$  with respect to centroid of coordinates ($\phi_h$, $\psi_k$).

 Notice that, even if Eq. (\ref{dalpha}) and Eq. (\ref{dbeta}) look mathematically identical, we prefer to  write down both formulas in order to point out that the differences may occur between the sets $\mathcal{I}$ and $\mathcal{P}$ and the related centroids coordinates. Indeed, as we will see below, the definition of the coordinates $\phi$'s and $\psi$'s and the different cardinalities of $\mathcal{I}$ and $\mathcal{P}$ lead to very different settings  emphasized in the cases  concerning Eq. (\ref{dalpha}) and Eq. (\ref{dbeta}).

Three cases of clustering search have been examined in (Ausloos et al., 2108a),  always setting $H=K=4$,  with the centroids regularly distributed on the plane diagonal: $\{\phi_h\}_{h=1}^H=\{\psi_k\}_{k=1}^K=\{1/5,2/5,3/5,4/5\}$. Consider  case [$II$], for discussion, when $\alpha_x=1/2$ for each $x \in \mathcal{I}$
and  an identical weight for the  $x \in \mathcal{P}$  variables, i.e., $\beta_x=1/7$. This is a  "\underline{uniform in value}" case, where the definition of the centroids is made by
considering a uniform decomposition of the interval $[0,1]$ and all
the variables are assumed to equally concur in the Voronoi distance.

It should be pointed out here that after some simple clustering analysis, so called  case  [$I$], in Ausloos et al. (2018a), 9 companies are  controlling the clustering,   and collapsing the whole sample into one single cluster, due to their "outlier aspect" They are : (2) Aeffe, (5)
Ascopiave, (15)
Cementir Holding, (20)
D'Amico, (22)
Digital Bros, (30)
Esprinet, (45)
Mondo Tv, (58)
Ternienergia, (59)
Tesmec.  This numerically confirms the few outlined cases seen in the above figures. These 9 companies are removed for the subsequent Voronoi clustering analysis approach.
 It remains therefore 53 companies to be examined.

To provide  comments on the following results, we call
\textit{first} cluster the one associated  with the
$\{\phi_h\}_{h=1}^H=\{\psi_k\}_{k=1}^K=\{1/5\}$ centroid and, in an
increasing way, the \textit{second} and the \textit{third} cluster,
so that the \textit{fourth} cluster is the one associated to the
values   $\{\phi_h\}_{h=1}^H=\{\psi_k\}_{k=1}^K=\{4/5\}$.

In Table \ref{Table3}, a description of the clusters of the sample companies is provided.

Referring to innovation clustering, the greatest number of companies
(45 out of 53) is located in the first cluster, meaning that, in
relative terms, companies undertake low value innovation initiatives
(at least those which produce reflections on tangible and intangible
assets). Total sales  and  total assets, which are measures largely
employed in literature for company size,  show that the higher the
intensity of innovation, the higher the size, or conversely.  Also
the incidence of both tangible and intangible assets (as percentage
of the total assets) is increasing in the three innovation clusters,
meaning that in highly innovative companies, tangible and intangible
assets represent a relevant portion of the disclosed  total assets.
The mean/std. dev. ratio shows that the composition of the clusters
is rather heterogeneous except for the 3rd innovation cluster which
is composed by companies whose size is fairly concentrated around
the mean. For what concern performance, the distribution of
companies among the clusters is quite different from that of
innovation.

 This provides evidence that the association between innovation and
performance is not self-evident. The averages in the performance
clusters also do not allow to appraise significant differences
neither in terms of company size or incidence of tangible and
intangible assets.

Table \ref{Table4} shows the averages of innovation and performance
drivers  for the entire sample and so called clustering $II$
analysis approach (Ausloos et al., 2018a) for innovation and
performance.

%\begin{center} NSERT TABLE 4 ABOUT HERE \end{center}
 For completeness, we reproduce comments from such a publication.
  In the first cluster, the averages of
innovation for tangible and intangible assets are below the general
averages  of  the entire sample, whereas all performance
indicators are above the  full sample averages. In the second cluster, a
general under-the-general-average performance is associated to an
above-the-general-average innovation. In the third cluster, the
performance averages are mixed: some of them are above the mean,
while the others are below.

Specifically, the $\mu/\sigma$ ratio  points that  the clusters
homogeneity   is  rather low, meaning that  extremely different
companies lie within the same cluster both from innovation or
performance perspective. The only exception is represented by ATO,
since the $\sigma$ is remarkably concentrated around the average
$\mu$. This could be interpreted as a possible association between
innovation and asset turnover. However, its direction remains
unclear, since a high  ATO  is associated to a low innovation in the
first cluster, whereas a low  ATO is associated to a medium
innovation in the second cluster,
 while     high  ATO  is associated to high innovation in the third cluster.

 It is worth noticing that in the third cluster,
companies appear rather homogeneous in terms of performance,
particularly for profitability (both ROI and ROS) and efficiency
(ATO and S/E).  One can  argue that,
above a particular  "threshold of innovation intensity", the level
performances seems rather homogeneous.

 Indeed, similar considerations can be
made for performance clustering: the performance averages gradually increase
from the first to the third cluster, whereas innovation averages
decrease (TIAX) or fluctuate (TTA). The
relation innovation-performance seems, then, quite puzzling. Even
for performance clustering, heterogeneity generally occurs within
the clusters except for  ATO.

\subsection{Evidence from outliers  approach}\label{Outtiers}

Since there is a negative minimum for each DS, DA, ROI and ROS, one
may guess that some board innovation strategies were rather
failures\footnote{In fact,  it is  not absolutely sure that innovation strategies were "failures", since one has no definite proof that innovations have been truly  implemented.}. One observes also some outliers from simple Voronoi
analysis; see case "Clustering I" in Ausloos et al. (2018a). From
Table \ref{Table1}, one observes that the kurtosis is always
positive and large, indicating lesser chances of extreme negative
outcomes.  The skewness is also positive, indicating   a long upper
tail (many small losses and a few extreme gains),  and a long lower
range tail (many small gains and several extreme losses).

The performance efficiency ratios of the (62) companies, taken one by one,   one observes several outliers, i.e. when the SME efficiency value fall outside the relevant   $]$$\mu-2\sigma$, $\mu+2\sigma$$[$   interval. There are 3 SMEs which are positive outliers: (58) Terrienergia, (11) Buongiorno, (13) Cairo Communications, and 1 SME which is systematically a "negative outlier": (45) Mondo TV, confirming the Voronoi analysis of "Clustering I".

It should occur to the reader that those 4 companies are those with
very low TTA.  Moreover, Mondo TV is the only one among the outliers
which has a TTA06  lower than its TTA07, - this SME had about a 50\%
decrease in investment before the crisis. In contrast, Terrienergia,
Buongiorno, and Cairo Communications have a relatively high TTA
increase.

One can observe  respectively from Fig. 12 and Fig. 13, for example; see also  Figs. 5-6 in Ausloos et al. (2018b).
 \begin{itemize}
\item      the relationship between $<$ROI$>$3 and $<$TTA$>$2: a weak  $<$ROI$>$3  for Cementir Holding and Ascopiave; a negative but with a  large  absolute  value occurs for D'Amico; in contrast, a large $<$ROI$>$3 occurs for Tesmec, while the negatively largest  $<$ROI$>$3   is for Eems, - both firms have a rather low $<$TTA$>$2;

\item    the relationship between $<$ROS$>$3 and  $<$TTA$>$2  indicates a moderate $<$ROS$>$3 positive effect  for Sogefi, Ascopiave, D'Amico and Cementir Holding, the four largest TTA companies, "imposing" a single cluster in  the "Clustering I" Voronoi analysis; a large negative $<$ROS$>$3 effect occurs for Mondo TV; on the other side, the most  positive  $<$ROS$>$3  is for Falck Renewables, Zignago Vetro, and Nice.
 \end{itemize}

Observe that these companies cover various sectors of activity.  Nevertheless, there are differences: Terrienergia and Cairo Communications have very dissimilar performance efficiency behaviors, the former performing better for "growth", the latter for "profitability". Since Terrienergia, Buongiorno, and Cairo Communications have a high increase in TTA, one might  reach some advice concerning innovation  strategy. Let TTA increase for better performance.

 As an {\it a posteriori"}  analysis "proof", observe that Mondo TV did not increase its TTA, TTA07$<$TTA06, pointing to a deficient strategy. This is pointing to the timing of "investment"  relevance, - not fully clear from the Voronoi analysis. Moreover, observe that   the average values are not the critical  quantities.

 \section{Conclusions} \label{Conclusions}

Justifying an investment  is superb challenge for board members., -
the more so at financial crisis time. Payback is unsure; one needs
criteria for  obtaining efficiency  (performance) measures, whence
modeling  strategies and forecasting.  Usually one demands that the
level of investments be low and the returns be high.  We have here
proposed a set of measures for such "research questions". We  have
outlined means for finding  correlations and looked for clustering
of performance ratios, whence the  specific companies "obeying"
strategies based on such ratios. But from the Voronoi clustering
approach,    it is not  obvious  that high innovation leads to high
performance.

We have found that the timing of investment is very relevant from
observing outliers, with either positive or negative results.
Extreme values show best strategies !

The Voronoi approach is nevertheless of interest: Within the
clusters, one can  compare the characteristics and performance of
companies holding the same innovation model, whereas between the
clusters high heterogeneity should occur assuming that different
innovation models might be suitable for different company profile
and/or could be associated to different level of performance.

In this respect, cluster analysis seems to be particularly effective
in providing a global analysis of the relationship between
innovation and performance. Notice that the study allows three
considerations from extreme value  analysis: not only the investment
evolution; up or down, low or high, but also through their average,
serving as a control kind of test. It should be obvious that the
best performance should be better appreciated when (unexpectedly?)
the investment is low, but regularly  implemented.

\clearpage

	\begin{table} \begin{center} 
\begin{tabular}[t]{|c|c|c||c|c|c|c} %ll 
  \hline
	$i=$& &	Supersector				&$i=$& &Supersector \\\hline					
	1	&	Acotel Group	&Telecommunications &	32	&	Exprivia	&	Technology\\		
	2	&	Aeffe	& P\&HG &	33	 &	Falck Renewables	&	Utilities	\\		
	3	&	Amplifon	&Health Care &	34	&	Fidia	&	IG\&S		\\
	4	&	Ansaldo Sts	& IG\&S &	35	&	Fiera Milano	& IG\&S\\			
	5	&	Ascopiave& Utilities	&	36	&	Gefran	&	IG\&S\\		
	6	&	Astaldi	&	C\&M &37	&	I.M.A	&	IG\&S		\\
	7	&	Biancamano	&  IG\&S &	38	&	Interpump Group	& IG\&S\\			
	8	&	Biesse	& IG\&S &	39	&	Irce	&		 IG\&S	\\
	9	&	Bolzoni	& A\&P	&40	&	Isagro	&	Chemicals	\\	
	10	&	Brembo	&	A\&P&41	&	It Way	&	Technology	\\	
	11	&	Buongiorno *	& Technology&	42	&	La Doria	&	F\&B\\		
	12	&	Cad It	&	Technology   &43	&	Landi Renzo	&A\&P	\\		
	13	&	Cairo Communic.&Media	&	44	&	Marr	& Retail\\			
	14	&	Cembre	&  IG\&S  &	45	&	Mondo Tv	&	Media	\\	
	15	&	Cementir Holding& C\&M 	&	46	&	Nice	&	IG\&S \\		
	16	&	Centrale Latte To&F\&B &	47	&	Panariagroup	& C\&M \\			
	17	&	Cobra	& Automobiles and Parts	&48	&	Poligraf. S. F	&	IG\&S	\\	
	18	&	Dada	& Technology	&49	&	Poltrona Frau	&	P\&HG	\\	
	19	&	Damiani	& Retail	&50	&	Prima Industrie		& IG\&S	\\	
	20	&	D'Amico	&  IG\&S &	51	&	Rdb			&	C\&M \\
	21	&	Datalogic	 &  IG\&S  &	52	&	Reno De Medici	& IG\&S\\			
	22	&	Digital Bros& P\&HG	&	53	&	Reply &Technology		\\		
	23	&	Dmail Group	& Media &	54	&	Sabaf	& IG\&S	\\		
	24	&	Dmt	&	Technology &55	&	Saes Getters		& IG\&S 	\\	
	25	&	Eems **	&Technology&	56	&	Servizi Italia	&  IG\&S	\\		
	26	&	El.En	&  IG\&S  &	57	&	Sogefi	&A\&P	\\	
	27	&	Elica	&  IG\&S  &	58	&	Ternienergia	&	Utilities		\\	
	28	&	Emak	& P\&HG  &	59	&	Tesmec	&	 IG\&S \\	
	29	&	Engineering&	Technology  &	60	&	Txt E-Solutions &Technology\\			
	30	&	Esprinet	&Technology&	61	&	Yoox	 & Retail ***\\
	31	&	Eurotech	&	Technology & 62	&	Zignago Vetro	&	 IG\&S\\			
  \hline 
\multicolumn{6}{|c|}{ *   Since July 2012, Buongiorno is part of Docomo Digital } \\  
\multicolumn{6}{|c|}{ **   Eems was moved  away from Technology in STAR  to MTA Market/Segment} \\  
\multicolumn{6}{|c|}{ *** In March 2015, Yoox merged with Net-a-Porter} \\  \hline 
 \end{tabular} 
   \caption{ The 62 STAR company names, at the time of study; alphabetical order and the "supersector" to which they belog; "supersector" abbreviations : Automobiles \& Parts (A\&P); Construction \& Materials (C\&M); Industrial Goods \& Services (IG\&S);  Personal \& Household Goods (P\&HG); Food \& Beverage	(F\&B)
     }\label{STARnames}
\end{center} \end{table}

 \clearpage
{\bf References}
  \vskip0.5cm
%\be{\bf gin{thebibliography}{00}

    \vskip0.5cm
Agostini, L., Nosella, A., and Soranzo, B. (2015).  The impact of formal
and informal appropriability regimes on SME profitability in medium
high-tech industries. Technology Analysis and Strategic Management
27(4), 405-419.

%Agostini, L., Nosella, A., \& Soranzo, B. (2015). The impact of formal and informal appropriability regimes on SME profitability in medium high-tech industries. Technology Analysis \& Strategic Management, 27(4), 405-419.

  \vskip0.5cm
Archibugi, D. (2001).  Pavitt's  taxonomy sixteen years on: a review
article, economics of innovation and new technology, Economics of
Innovation and New Technology 10(5), 415-425.

  \vskip0.5cm
 Ausloos, M.,  Bartolacci, F., Castellano, N.G.,  and Cerqueti, R.  (2018).  Exploring how innovation strategies at time of crisis influence performance: a cluster analysis perspective, Technology Analysis and Strategic Management, 30(4), 484-497.

  \vskip0.5cm
 Ausloos, M., Cerqueti, R.,  Bartolacci, F., and Castellano, N.G. (2018). 
 SME investment best strategies. Outliers for assessing how to optimize performance,
 Physica A 509, 754-765.

  \vskip0.5cm
Archibugi, D., Filippetti, A., and Frenz, M. (2013).  Economic crisis and innovation: Is destruction prevailing over accumulation? Research
Policy 42(2), 303-314.

  \vskip0.5cm
Bartolacci, F., Castellano, N.G., and   Cerqueti, R. (2015).  The impact of
innovation on companies'performance: an entropy-based analysis of the STAR Market Segment of Italian Stock Exchange, Technology
Analysis and Strategic Management 27(1), 102-123.

  \vskip0.5cm
Baum, C.F., L\"{o}\"{o}f, H., Nabavi, P., and Stephan, A. (2017).  A new
approach to estimation of the $R\& D$-innovation-productivity relationship. Economics of Innovation and New Technology 26(1/2),
121-133.

  \vskip0.5cm
 Bartolacci, F.,   Castellano, N.G.,  and Cerqueti, R. (2015).   The impact of innovation on companies' performance: an entropy-based analysis of the STAR Market Segment of Italian Stock Exchange, Technology Analysis and Strategic Management 27(1), 102-123.

  \vskip0.5cm
Brusoni, S., Cefis, E., and Orsenigo, L. (2006).  Innovate or Die? A
critical review of the literature on innovation and performance,
CESPRI, Centro di Ricerca sui Processi di Innovazione e
Internazionalizzazione, Universit\'{a} Commerciale  Luigi Bocconi,
Working Paper n. 179.

  \vskip0.5cm
Ceptureanu, E.G., Ceptureanu, S.I., Luchian, C.E., and  Luchian. I. (2017).  Quality Management in Project Management Consulting. A Case Study in an International Consulting Company. Amfiteatru Economic 19(44), 215-230.

  \vskip0.5cm
Cesaratto, S., and Mangano, S. (1993).  Technological profiles and economic performance in the Italian manufacturing sector. Economics of Innovation and New Technology 2, 237-256.

%  \vskip0.5cm
%Choi, S.B., Williams, C. (2016).  Entrepreneurial orientation and
%performance: mediating effects of technology and marketing action
%across industry types. Industry and innovation 23(8), 673-693.

  \vskip0.5cm
Chun, D., Chung, Y., and Bang, S. (2015).  Impact of firm size and
industry type on $R\& D $ efficiency throughout innovation and commercialisation stages: evidence from Korean manufacturing firms.
Technology Analysis and Strategic Management 27(8), 895-909.

  \vskip0.5cm
Cooper, R.G. (1984). The strategy-performance link in product innovation. R\& D Management 14(4), 247-259.

  \vskip0.5cm
De Jong, J.P.J., and  Marsili, O. (2006).  The fruit flies of innovations: A taxonomy of innovative small firms. Research Policy 35(2),
213-229.

\vskip0.5 cm
Descartes, R. (1644). Principia Philosophiae. Ludovicus Elzevirius, Amsterdam, 1644

  \vskip0.5cm
Duyckaerts, C., and Godefroy, G. (2000). Voronoi tessellation to study the numerical density and the spatial distribution of neurones.
Journal of Chemical Neuroanatomy 20(1), 83-92.

  \vskip0.5cm
Dwyer, L., and Mellor, R. (1993).  Product innovation strategies and performance of Australian firms. Australian Journal of Management
18(2), 159-180.

  \vskip0.5cm
Filippetti, A., and Archibugi, D. (2011).  Innovation in times of crisis:
National Systems of Innovation, structure, and demand. Research
Policy 40(2), 179-192.

  \vskip0.5cm
Gadomski, A. and Kruszewska, N. (2012).  On clean grain-boundaries
involving growth of nonequilibrium crystalline-amorphous superconducting materials addressed by a phenomenological viewpoint.
The European Physical Journal B 85(12), 1-8.

  \vskip0.5cm
Gligor, M., and Ausloos, M. (2007).  Cluster structure of EU-15 countries
derived from the correlation matrix analysis of macroeconomic index fluctuations, The European Physical Journal B 57, 139-146.

  \vskip0.5cm
Gligor, M., and Ausloos, M. (2008a).  Convergence and cluster structures
in EU area according to fluctuations in macroeconomic indices,
Journal of Economic Integration 23, 297-330.

  \vskip0.5cm
Gligor, M., and Ausloos, M. (2008b).  Clusters in weighted macroeconomic
networks: the EU case. Introducing the overlapping index of GDP/capita fluctuation correlations, The European Physical Journal B
63, 533-539.

  \vskip0.5cm
Gocer, I., Alatas, S., and Peker, O. (2016).  Effects of $R\& D$ and
innovation on income in EU countries: new generation panel cointegration and causality analysis. Theoretical and Applied
Economics 23(4), 153-164.

  \vskip0.5cm
Heirman, A., and Clarysse, B. (2007).  Which tangible and intangible assets matter for innovation speed in start-ups? Journal of Product
Innovation Management 24(4), 303-315.

  \vskip0.5cm
Hitt, M.A., Hoskisson, R.E., and Kim, H. (1997). International diversification: effects on innovation and firm performance in
product diversified firms. Academy of Management Journal 40(4), 767-798.

  \vskip0.5cm
Hollenstein, H. (2003). Innovation modes in the Swiss service sector: a cluster analysis based on firm-level data. Research Policy 32,
845-863.

  \vskip0.5cm
Jensen, M.B., Johnson, B., Lorenz, E., and Lundvall, B.A. (2007).  Forms
of knowledge and modes of innovation. Research Policy 36, 680-693.

%  \vskip0.5cm
%Kafouros, M.I., Wang, C., 2008 The Role of Time in Assessing the
%Economic Effects of R\&D. Industry and innovation 15(3), 233-251.

  \vskip0.5cm
Khan, A.M., and Manopichetwattana, V. (1989). Innovative and non-innovative small firms: types and characteristics. Management
Science 15(5), 597-606.

  \vskip0.5cm
Kirner, E., Kinkel, S., and Jaeger, A. (2009).  Innovation paths and the innovation performance of low-technology firms - An empirical analysis
of German industry. Research Policy 38, 447-458.

  \vskip0.5cm
Latham, S.   and Braun, M. (2011). 
 Economic recessions, strategy, and performance: a synthesis  Journal of Strategy and Management 4 (2), 96-115.

  \vskip0.5cm
Lawless, M.W. and Anderson, P.C. (1996). Generational technological
change: effects of innovation and local rivalry on performance. Academy of Management Journal 39(5), 1185-1217.

  \vskip0.5cm
Leiponen, A. and Drejer, I. (2007).  What exactly are technological regimes? Intra-industry heterogeneity in the organization of
innovation activities. Research Policy 36, 1221-1238.

  \vskip0.5cm
Liu, X.T., Zheng, X.Q., and Li, D.B. (2009).  Voronoi Diagram-Based Research on Spatial Distribution Characteristics of Rural
Settlements and Its Affecting Factors. A Case Study of Changping District, Beijing [J]. Journal of Ecology and Rural Environment 2, 007.

  \vskip0.5cm
Montresor, S. and Vezzani, A. (2016).  Intangible investments and
innovation propensity: Evidence from the Innobarometer 2013.
Industry and Innovation 23(4), 331-352.

%Montresor, S., \& Vezzani, A. (2016). Intangible investments and innovation propensity: Evidence from the Innobarometer 2013. Industry and Innovation, 23(4), 331-352.

  \vskip0.5cm
Nunes, S. and Lopes, R. (2015).  Firm performance, innovation modes and
territorial embeddedness, European Planning Studies 23(9),
1796-1826.

  \vskip0.5cm
OECD, (2005).  Annual Report, Paris: OECD Publishing.

  \vskip0.5cm
OECD, (2009).  Annual Report, Paris: OECD Publishing.

  \vskip0.5cm
Park, N.K., Park, U.D., and Lee, J. (2012).  Do the Performances of
Innovative Firms Differ Depending on Market-oriented or
Technology-oriented Strategies?. Industry and Innovation 19(5),
391-414.

  \vskip0.5cm
Pavitt, K. (1984). Sectoral patterns of technical change: towards a
taxonomy and a theory. Research Policy, 13, 343-373.

  \vskip0.5cm
Ramella, M., Boschin, W., Fadda, D., and Nonino, M. (2001).  Finding
galaxy clusters using Voronoi tessellations. Astronomy \&
Astrophysics 368(3), 776-786.

  \vskip0.5cm
Ranga, M. and  Etzkowitz, H. (2012).  Great expectations: an innovation
solution to the contemporary economic crisis. European Planning
Studies 20(9), 1429-1438.

  \vskip0.5cm
Renzi, A. and  Simone, C. (2011).  Innovation, tangible and intangible
resources: The espace of slacks interaction. Strategic Change
20(1-2), 59-71.

  \vskip0.5cm
Shin, N., Kraemer, K.L., and Dedrick J. (2017).  R\&D and firm performance
in the semiconductor industry. Industry and Innovation 24(3),
280-297.

  \vskip0.5cm
Srholec, M. and  Verspagen, B. (2012).  The Voyage of the Beagle into
innovation: Explorations on heterogeneity, selection, and sectors.
Industrial and Corporate Change 21(5), 1221-1253.

  \vskip0.5cm
Sterlacchini A. and  Venturini F. (2014).  R\&D and Productivity in
High-Tech Manufacturing: A Comparison between Italy and Spain.
Industry and Innovation 21(5), 359-379.

  \vskip0.5cm
Tseng, C.Y., Hui-Yueh Kuo, H-Y., and Chou, S.S. (2008).  Configuration of
innovation and performance in the service industry: evidence from
the Taiwanese hotel industry. The Service Industries Journal. 28(7),
1015-1028.

  \vskip0.5cm
Vaz, E., de Noronha Vaz, T., Galindo, P.V., and Nijkamp, P. (2014). 
Modelling innovation support systems for regional
development analysis of cluster structures in innovation in
Portugal. Entrepreneurship and Regional Development 26(1-2), 23-46.

  \vskip0.5cm
Voronoi, G.F. (1908). Nouvelles applications des param\'{e}tres
continus de la th\'{e}orie de formes quadratiques, Journal f\"{u}r
die reine und angewandte Mathematik 134, 198-287.

  \vskip0.5cm
Yushimito, W.F., Jaller, M., and Ukkusuri, S. (2012).  A Voronoi-based
heuristic algorithm for locating distribution centers in disasters.
Networks and Spatial Economics 12(1), 21-39.

  \vskip0.5cm
Zahra, S.A. and  Covin, J.G. (1994). Domestic and international
competitive focus, technology strategy and company performance: an
empirical analysis. Technology Analysis and Strategic Management
6(1), 39-54.

%\end{thebibliography}

\clearpage
 
 \begin{table}% \begin{center}
\begin{tabular}[t]{|c| c|c|c|c|c|c| c|c|c|}
  \hline
& \multicolumn{2}{|c|}{   Innovation}      &       \multicolumn{6}{|c|}{   Performance}\\\hline
    &       \multicolumn{2}{|c|}{    }       &       \multicolumn{2}{|c|}{   Growth }  &       \multicolumn{2}{|c|}{   Profitability}&     \multicolumn{2}{|c|}{  Efficiency}\\\hline
 &   Intangible     &   Tangible    &   Sales   &   $Tot.Ass.$    &   $Ret.on$      &$Ret.on$     &   Asset & Sales/  \\
  &  Assets &   Assets      &    $Var.n$  &     $Var.n$ &  Invest.  &   Sales   &turnover     & $empl.$\\
   & (TIAX)     &    (TTA) &      (DS)   &    (DA)  &   (ROI)   &  (ROS)    & (ATO)&  (S/E)\\\hline
mean ($\mu$)&12,360.46  &   29,215.40   &   9\% &   6\% &   5\% &   5\% &   0,91    &   275.77  \\
Std.Dev.($\sigma$)&18,695.11    &   45,379.80   & 16\%  &   14\%    &   5\% &   7\% &   0,34    &   231.20  \\
$\mu/\sigma$&0.66   &   0.64    &   0,46    &   0,57    & 0,85  &   0,75    &   2,68    &   1.19    \\
  min.&  180    &   86.50   &   -19\%   &   -10\%       &   -8\%    &   -14\%   &   0,15    &   57.20   \\
Max&80,816  &     217,237.50    &   59\%    &   53\%    &   21\%    &   24\%    &   2,04    &   1,100.76    \\
Q1&1,346.50     &       3,579.50    &   -1\%    &   -2\%    &   2\% &   1\% &   0,75    &   148.02  \\
median&  3,584  &     10,329.00     &   3\% &   4\% &   4\% &   5\% &   0,86    &   188.04  \\
Q3&13,917   &   31,331.50   &   12\%    &   16\%    &   8\% &   9\% &   1,09    &   281.07  \\
Skewness&2.28   &   2.57    &   1.48    &   1.32    &       0.44    &   0.27    &   0.78    &   2.25    \\
 Kurtosis&      5.14    &              6.79     &   3.60    &   1.07    &   0.60    &   0.76    &   1.76    &   5.05    \\ \hline
\end{tabular}
 \caption{  Main statistical indicators of the innovation and performance variables; $Tot.Ass.$ = "Total Assets"; $Var.n$ = "variation"; $empl.$ = "employee"; %$N.$ = "Number of"; 
 $Ret.on$= returns on .} \label{Table1}
%\end{center}
 \end{table}

   \begin{table} \begin{center}
\begin{tabular}[t]{|c||c|c|c|c|c|c|c|}
  \hline
%& \multicolumn{2}{|c|}{ $ATI_{c,p}$}&  \multicolumn{2}{|c|}{ $N_{inhab,c,p}$}  \\ \hline

Variable    &   Min.    &   Max.    &   Sum &   Mean    &   StDev   &   Skewness    &   Kurtosis    \\
    &       &       &       &   ($\mu$) &   ($\sigma$)  &       &           \\\hline
$<$TIAX$>$2  &   174.5   &   1.192   10$^5$ &     8.421 10$^6$       &   13 583  &   22 513  &   2.7259  &   8.0364  \\
$<$TTA$>$2   &   86.5    &   5.075   10$^5$  &   2.746 10$^6$    &   44 297  &   92 600  &   3.3967  &   12.062  \\\hline
$<$DS$>$3    &   -0.1924 &   1.1767  &   4.9303  &   0.0795  &   0.198   &   3.1414  &   14.013  \\
$<$DA$>$3&   -0.1436 &   1.9818  &   7.8786  &   0.1270  &   0.330   &   3.8060  &   16.885  \\
$<$ROI$>$3&  -0.0768 &   0.3457  &   3.0115  &   0.0486  &   0.067   &   1.5342  &   5.1206  \\
$<$ROS$>$3&  -0.6609 &   0.2445  &   2.5316  &   0.0408  &   0.118   &   -3.505  &   20.046  \\
$<$ATO$>$3&   0.1474&    3.5673  &59.900& 0.9661  &   0.535  &   2.4625  &8.8557     \\
$<$S/E$>$3&      17.464  &787.69 &   7739.5  &   124.83 & 155.6& 2.9856 &    8.7591      \\
\hline
 \end{tabular}
   \caption{Summary of  (rounded) statistical characteristics  for  the time average distributions of the innovation  and performance indicators for the 62 STAR companies,  in the center of the table, in per cent and in $10^6$ Euros, respectively; the skewness and kurtosis are dimensionless scalars.
   }\label{TablestatSKSK}
\end{center} \end{table}

\clearpage

 \begin{table} \begin{center}
\begin{tabular}[t]{|c| c|c|c|c|c|c| }
  \hline
% \multicolumn{7}{|c|}{  II clustering  }   \\  \hline
 \multicolumn{2}{|c} { "II clustering"  }&   \multicolumn{5}{|c|}{  Performance }     \\ \hline
\multicolumn{1}{|c} {   }&& $1st$ $cl.$&    $2nd$ $cl.$ &   $3rd$ $cl.$ &   $4th$ $cl.$&    Tot.     \\ \hline
Innovation  &   $1st$ $cl.$ &   16  &   22  &   7   &   0   &   45      \\
    &   $2nd$ $cl.$ &   2   &   2   &   0   &   0   &   4    \\
    &   $3rd$ $cl.$ &   1   &   3   &   0   &   0   &   4       \\
    &   $4th$ $cl.$&0   &   0   &   0   &   0   &   0   \\   \hline
     \multicolumn{2}{|c|}   {Tot.}  &   19  &   27  &   7   &   0   &   53    \\ \hline
  \end{tabular}
 \caption{  Distribution   of companies among the clusters ($cl.$), either for clustering $II$, as defined in the text and examined in Ausloos et al. (2018a) for 53 STAR companies } \label{Table2}
\end{center} \end{table}

  \clearpage

   \begin{table} \begin{center}
\begin{tabular}[t]{|c|c| ccccc|}
  \hline
     \multicolumn{3}{|c|}{}     &   $Tot.Ass.$  &   Total    & Intangible    &    Tangible     \\
  \multicolumn{3}{|c|}{ }    &  2006-2007   &   Sales &  Assets on  &    Assets on      \\
  \multicolumn{3}{|c|}{     }       &    (\EUR/1,000)   &    (\euro/1,000)  &    $Tot.Ass.$ &  $Tot.Ass.$   \\ \hline
    & N:    &   Mean    &   303,053     &   267,689     &   5\% &   10\%    \\
&    53 &       Std.Dev.   &   304,144     &   261,828     &   6\% &   12\%    \\ \hline \hline
    \multicolumn{6}{|c}{  "II clustering"  Innovation } &\\ \hline
    &           &   Mean    &   241,199     &   210,452         &   4\% &   8\% \\
    $1st$ $cl.$ &   45  &   Std.Dev.   &   190,099     &   187,814         &   5\% &   8\% \\
        &   &   Mean/St.Dev.   &   1.27    &   1.12    &       0.88    &   1.04    \\ \hline
 &      &   Mean    &   731,745     &   467,442         &   8\% &   16\%    \\
 $2nd$ $cl.$    &   4   &   Std.Dev.   &   798,924     &   469,023         &   8\% &   22\%    \\
 &      &   Mean/St.Dev.   &   0.92    &   1.00    &       0.97    &   0.73    \\ \hline
 &      &   Mean    &   570,226     &   711,853         &   12\%    &   22\%    \\
$3rd$  $cl.$  &   4   &   Std.Dev.   &   261,414     &   329,233         &   9\% &   27\%    \\
    &       &       Mean/St.Dev.   &   2.18    &   2.16    &   1.39    &   0.82    \\ \hline\hline

    \multicolumn{6}{|c}{  "II clustering"  Performance   } &\\ \hline
    &        &  Mean    &   210,607     &   157,375     &       8\% &   10\%    \\
 $1st$ $cl.$    &   19  &   Std. Dev.   &   130,310     &   118,218         &   7\% &   11\%    \\
    &       &   Mean/St.Dev.   &   1.62    &   1.33    &       1.14    &   0.94    \\ \hline
&       &   Mean    &   385,628     &   354,434         &   3\% &   9\% \\
 $2nd$ $cl.$    &   27  &   Std.Dev.   &   387,615     &   293,834         &   3\% &   13\%    \\
    &       &       Mean/St.Dev.   &   0.99    &   1.21    &       0.90    &   0.71    \\ \hline
    &           &   Mean    &   235,476     &   232,525         &   3\% &   10\%    \\
   $3rd$  $cl.$  &   7   &   Std.Dev.   &   228,112     &   340,089         &   5\% &   12\%    \\
    &       &       Mean/St.Dev.   &   1.03    &   0.68    &       0.62    &   0.85        \\ \hline
\end{tabular}
 \caption{ Statistical description of the  companies, as if one full sample, or "belonging" to a cluster $cl.$   (see text); the number (N)  of companies in each cluster is given;    $Tot.Ass.$ = "Total Assets".} \label{Table3}
\end{center} \end{table}
  \clearpage

   \begin{table} \begin{center}
\begin{tabular}[t]{|cc| cccccccc|}
  \hline
&    \multicolumn{2}{c}{}  TIAX &   TTA &   DS   &   DA  &  ROI    &   ROS &   ATO     &   S/E\\
% &       \multicolumn{2}{c}{}  &     &  \% &    \%     &       &       &  &        \\
  \hline \hline
Ent.&   Mean ($\mu$)    &     12,360    &     29,215    &   6\% &   9\% &   5\% &   5\% &   0.91    &   275.77  \\
 &  Std.Dev.($\sigma$)  &     18,695    &     45,380    &   14\%    &   16\%    &   5\% &   7\% &   0.34    &   231.20  \\
 N=53&  $\mu/\sigma$    &   0.66    &   0.64    &   0.46    &   0.57    &       0.85    &      0.75     &     2.68  &      1.19     \\  \hline\hline
    \multicolumn{6}{|c}{Innovation II clustering} \\ \hline
$1st$ &Mean ($\mu$)           &   7,127   &   18,554  &   7\% &   9\%     &   5\% &   6\% &   0.93    &   293.83  \\
$cl.$ &   Std.Dev.($\sigma$)  &     9,311     &   24,023  &   15\%    &   17\%    &   6\% &   7\% &   0.36    &   248.53  \\
N=45    &   $\mu/\sigma$        &   0.77        &   0.77        &   0.49    &   0.55    &   0.87    &    0.79 & 2.57    &   1.18    \\  \hline
$2nd$  &  Mean ($\mu$)        &   28,081  &   71,512  &   4\% &   3\% &   2\% &   2\% &   0.67    &   165.90  \\
$cl.$ &   Std.Dev.($\sigma$)  &   29,505  &   65,192  &   11\%    &   7\% &   6\% &   10\%    &   0.13    &   58.91   \\
N=4 &   $\mu/\sigma$        &   0.95        &   1.10        &   0.34    &   0.35    &   0.25    &   0.16    &   5.36    &   2.81    \\  \hline
$3rd$ &  Mean ($\mu$)        &   55,511  &   106,862     &   1\%     &   14\%    &   4\% &   5\% &   0.97    &   182.49  \\
$cl.$ &   Std.Dev.($\sigma$)  &   28,454  &   107,418     &   5\%     &   18\%    &   1\% &   2\% &   0.20    &   48.84   \\
N=4 &   $\mu/\sigma$        &   1.95        &   0.99        &   0.19    &   0.81    &   3.00    &   2.57    &   4.91    &   3.73    \\  \hline\hline
    \multicolumn{6}{|c}{Performance II clustering} \\ \hline
$1st$   & Mean ($\mu$)        &   16,356  &    22,484     &   -2\%    &   -4\%    &   0\% &   0\% &   0.79    &   222.75  \\
 $cl.$    &   Std.Dev.($\sigma$)  &    20,762     &   31,548  &   10\%    &   4\% &   4\% &   7\% &   0.28    &   157.65  \\
N=19&   $\mu/\sigma$        &       0.79    &   0.71        &   0.21    &   0.83    &   0.05    &    0.05 & 2.85    &   1.41    \\  \hline
$2nd$  &  Mean ($\mu$)        &    11,930     &     35,841    &   9\% &   12\%    &   7\% &   8\% &   0.94    &   262.81  \\
$cl.$ &   Std.Dev.($\sigma$)  &    19,346     &   56,259  &   9\% &   13\%    &   4\% &   5\% &   0.25    &   232.07  \\
N=27&   $\mu/\sigma$        &       0.62    &   0.64        &   0.94    &   0.92    &   1.54    &   1.46    &   3.77    &   1.13    \\  \hline
$3rd$ &  Mean ($\mu$)        &   3,174   &    21,931     &   20\%    &   35\%    &   9\% &   12\%    &   1.11    &   469.70  \\
$cl.$ &   Std.Dev.($\sigma$)  &      4,744    &    32,962     &   24\%    &   17\%    &   6\% &   9\% &   0.65    &   332.70  \\
N=7 &   $\mu/\sigma$        &   0.67        &   0.67        &   0.84    &   2.02    &   1.49    &   1.26    &   1.71    &   1.41    \\   \hline
\end{tabular}    \caption{ Main statistical characteristics of the Innovation and Performance  variables for the whole sample (Ent.) or inside the clusters ($cl.$); the number (N) of companies inside each cluster is recalled.} \label{Table4}
\end{center} \end{table}

\end{document}